\documentclass[journal]{IEEEtran}

\usepackage[utf8x]{inputenc}
\usepackage{tikz,pgfplots}
\usepackage{url}
\usepackage{graphicx}
\usepackage{subfig}
\usepackage{amsmath}
\usepackage{amsthm}
\usepackage{booktabs}
\usepackage{listings}
\usepackage{multirow}
\usepackage{algorithm}
\usepackage{algorithmic}

\theoremstyle{plain}

\usetikzlibrary{arrows,automata,calc,shapes.symbols}
\usetikzlibrary{positioning}
\usetikzlibrary{fit}
\usetikzlibrary{plotmarks}

\tikzset{
    >=stealth',
    operation/.style={
           rectangle,
           rounded corners,
           draw=black,
           text width=1.25cm,
           minimum height=0.5cm,
           text centered},
    operation-wide/.style={
           rectangle,
           rounded corners,
           draw=black,
           text width=1.7cm,
           minimum height=0.8cm,
           text centered},
    rect/.style={
           rectangle,
           draw=black, thick,
           minimum height=1em,
           text centered},
    input/.style={
           minimum height=1em,
           text centered},
    line/.style = {draw, ->},
    narline/.style = {text width=2cm, font = \small}
}

\title{Correction Trees as an Alternative to\\ Turbo Codes and Low Density Parity Check Codes}

\author{
  Jarosław Duda, Paweł Korus,~\IEEEmembership{Student Member,~IEEE,}
  \thanks{J. Duda is with the Institute of Physics, Jagiellonian University, ul. Reymonta 4, 30-059 Krak\'{o}w, Poland. E-mail:  dudaj@interia.pl}
  \thanks{P. Korus is with the Department of Telecommunications, AGH University of Science and Technology, al. Mickiewicza 30, 30-059 Krak\'{o}w, Poland. E-mail: pkorus@agh.edu.pl}
  \thanks{The research leading to these results has received funding from the INDECT project funded by European Community's Seventh Framework Programme (FP7 / 2007-2013) under grant agreement no.[218086].}
}

\newcommand{\bea}{\begin{eqnarray}}
\newcommand{\eea}{\end{eqnarray}}
\newcommand{\be}{\begin{equation}}
\newcommand{\ee}{\end{equation}}
\newcommand{\beas}{\begin{eqnarray*}}
\newcommand{\eeas}{\end{eqnarray*}}

\newcommand{\bc}{\begin{center}}
\newcommand{\ec}{\end{center}}

\newcommand{\nn}{n}
\newcommand{\kk}{k}
\newcommand{\RR}{R}
\newcommand{\pb}{\epsilon}
\newcommand{\ttemp}{z}
\newcommand{\rmsk}{r_m}

\newcounter{MYtempeqncnt}

\begin{document}

\maketitle

\begin{abstract}
The rapidly improving performance of modern hardware renders convolutional codes obsolete, and allows for the practical implementation of more sophisticated correction codes such as low density parity check (LDPC) and turbo codes (TC). Both are decoded by iterative algorithms, which require a disproportional computational effort for low channel noise. They are also unable to correct higher noise levels, still below the Shannon theoretical limit. In this paper, we discuss an enhanced version of a convolutional-like decoding paradigm which adopts very large spaces of possible system states, of the order of $2^{64}$. Under such conditions, the traditional convolution operation is rendered useless and needs to be replaced by a carefully designed state transition procedure. The size of the system state space completely changes the correction philosophy, as state collisions are virtually impossible and the decoding procedure becomes a correction tree. The proposed decoding algorithm is practically cost-free for low channel noise. As the channel noise approaches the Shannon limit, it is still possible to perform correction, although its cost increases to infinity. In many applications, the implemented decoder can essentially outperform both LDPC and TC. This paper describes the proposed correction paradigm and theoretically analyzes the asymptotic correction performance. The considered encoder and decoder were verified experimentally for the binary symmetric channel. The correction process remains practically cost-free for channel error rates below 0.05 and 0.13 for the 1/2 and 1/4 rate codes, respectively. For the considered resource limit, the output bit error rates reach the order of $10^{-3}$ for channel error rates 0.08 and 0.18. The proposed correction paradigm can be easily extended to other communication channels; the appropriate generalizations are also discussed in this study.
\end{abstract}


\section{Introduction}

Approaching the Shannon limit in forward error correction inherently involves operating on data blocks of rapidly increasing size. Alongside this growth, the number of valid codewords increases exponentially. As a result, the correction process, which involves finding the closest valid codeword to the received one, quickly becomes impractical. Error correction methods are considered practical if they are able to find an approximation of the optimal correction in reasonable time. In general, there are two main approaches to the problem \cite{mackay2003}. The first is to spread pseudo-random verification bits uniformly over the transmitted message. This approach is adopted by convolutional codes \cite{conv} and their derivatives, such as turbo codes (TC) \cite{berrou}. The second approach uses linear block codes with very low degrees of vertices. Such codes are known as low density parity check (LDPC) or Gallager codes \cite{gallager}.

Both LDPC and TC have their limitations. The former are defined by very large sparse matrices, whose generation and control during the correction process requires significant computational resources. Both LDPC and TC are decoded by iterative algorithms, usually with a pre-defined number of iterations. As such, they are characterized by a nearly constant decoding cost, regardless of the noise level actually observed in the channel. Due to the increasing capabilities of modern hardware decoders, this cost is now becoming acceptable, and the methods emerge as a replacement for conventional convolutional codes.

The family of convolutional codes is in practice limited to relatively small spaces of possible system states, which stems from the inconveniences caused by the intrinsic convolution operation. The typical size of the system state space is at most $2^{16}$. As a result, wrong correction patterns frequently generate correct system states, and instead of a complete correction, the decoder is merely able to reduce the number of errors in the message. The TC addressed this issue by repeating the encoding process on an interleaved version of the data blocks. They use half of the redundancy in every step, and require to consider the entire system state history. Hence, the correction of even low error rates already requires relatively long times. For convolutional codes, there is a computational cut-off rate \cite{cutoff} above which the expected correction time per symbol grows to infinity for infinite data streams. We use finite fixed length data frames (mainly 1kB), which enables efficient operation with rates up to the Shannon limit.

In this study, we analyze the correction potential of adopting large spaces of possible system states in a convolutional-like coding process. We will show that such an approach leads to very efficient correction algorithms, which enable correction arbitrarily close to the Shannon limit. We consider the space of at least $2^{64}$ system states, where the probability of system state collision for a wrong correction pattern becomes negligible ($2^{-64}\approx 5\cdot 10^{-20}$). This approach becomes tractable by designing a decoder which processes a very small fraction of the state space. We will show that the decoding process can be made practically cost-free for small error rates, and that no additional correction limits arise. Theoretically, a once encoded message can always be nearly completely corrected under the Shannon limit; however, in this limit the decoding cost grows to infinity.

Due to the negligible system state collision probability, the correction structure practically no longer contains cycles, and becomes a tree instead. The nodes of this tree represent a correction path up to a given position in the data stream. This approach imposes sequential decoding, which considers the most probable branch of the tree in each decoding step. The probability of a tree branch is reflected in the \emph{weights} of the corresponding tree nodes. The greater the weight, the more probable the corresponding node. For the purpose of efficient access to successive prospective correction paths, we use a heap to store the nodes of the tree. The heap operations have logarithmic complexity, which constitutes a significant improvement over commonly used linear-complexity structures, such as the stack.

For large error concentrations, the proposed correction algorithm may require impractically large numbers of necessary decoding steps. For any acceptable total number of decoding steps, there is a nonzero probability that the given step limit is insufficient. We refer to such situations as \emph{critical error concentrations} (CEC). In this study, we provide a theoretical methodology for quantitative analysis of the error concentrations, their probability distributions, and the expected number of decoding steps in which the decoder can deal with them. We will focus on the binary symmetric channel (BSC) and discuss different channel noises.

Susceptibility to critical error concentrations stems from the sequential nature of data stream processing. In traditional convolutional codes, where the state space is strongly limited, this problem can be mitigated by guessing the correct state after a concentration of errors. To some extent, error bursts in the channel can be dispersed with the adoption of stream interleavers. In order to address this issue, we adopt two techniques. Firstly, we construct a systematic code in order to bound the output error rate by the input error rate. Secondly, we design the scheme so that it would be possible to process the stream from both the beginning and the end. After reaching the same place in the stream, the correction paths are merged. With bidirectional correction, a data block will remain essentially damaged if there are at least two independent CECs. The probability of such an event is approximately equal to the square of the probability of a single CEC.

In summary, the main contribution of our work addresses the following issues. Firstly, we consider a much larger system state space and design of a tailored fast coding and decoding scheme. Secondly, we adopt a logarithmic-time heap structure instead of a commonly used linear-time stack. Thirdly, we propose an efficient state transition procedure, optimized for both forward and backward processing. We also design a dedicated mechanism, based on look-up tables, which enables rapid pruning of invalid tree branches. The decoder considers just the correction patterns which are guaranteed to lead to the allowed system states.

The paper is organized as follows. Section~\ref{sec:algorithm} describes the proposed encoding and decoding algorithms. Theoretical analysis of the decoding performance and the impact of error concentrations is presented in Section~\ref{sec:theory}. The results of experimental evaluation are shown in Section~\ref{sec:evaluation}. Further perspectives for adapting to different communication channels are briefly presented in Section~\ref{sec:perspectives}. A discussion of the applicability of the proposed scheme and final conclusions are presented in Section~\ref{sec:conclusions}.

\section{The Correction-Tree Algorithm}
\label{sec:algorithm}

This section describes the principles of the proposed correction algorithm. We begin with an explanation of the encoding procedure and a corresponding forward correction algorithm. We then extend the algorithm to a bidirectional case. A theoretical foundation for the utilized error correction approach can be found in Section~9 of \cite{ans}.

The following notation is used in the paper:\\

\begin{tabular}{rp{7cm}}
  $\propto$ &  proportional to \\
  $\oplus$ & bit-wise exclusive disjunction \\
  $\&$ & bit-wise logical conjunction \\
  $\tilde{p}$ & $=1-p$ \\
  $\lg$ & $\log_2$\\
  $h(p)$ & $=-p\lg(p)-\tilde{p}\lg(\tilde{p})$\\
  \#$A$ & cardinality of set $A$\\
  $\nn$ & total number of bits in encoded symbols \\
  $\kk$ & number of payload bits in encoded symbols \\
  $\RR$ & $=\nn-\kk$ - the number of redundancy bits \\
  $\kk/\nn$ & code rate (excluding the final state) \\
  $p_d$ & $=1-2^{-\RR}$ - probability that redundancy bits would accidentally agree \\
  $l$ & number of symbols in data frame \\
  $\pb$ & probability of bit flip in a binary symmetric channel (BSC) \\
  $p_d^0$ & $=1-2^{-\nn h(\pb)}$ - BSC Shannon limit for $p_d$ \\
  $cn$ & current node of the correction tree \\
  $cn.d$ & data symbol position associated with $cn$ \\
  $cn.p$ & parent node of $cn$ \\
  $cn.s$ & system state associated with $cn$ \\
  $cn.e$ & correction pattern associated with $cn$ \\
\end{tabular}

\subsection{Encoding Procedure}

The operation of the encoder begins with initializing the system state and the state transition tables. The algorithm operates sequentially on successive symbols from the data stream. Upon reception of a new symbol, the encoder performs symbol-dependent transition of the system state and extracts the redundancy as a certain number of the least significant bits of a binary representation of the posterior system state. This redundancy is transmitted together with the symbol in a systematic manner, i.e., by concatenation of the payload and the redundancy information.

There are two important consequences of this approach. Firstly, the transmitted stream is equipped with uniformly distributed redundancy information. Secondly, the redundancy of all transmitted symbols is intuitively connected by the system state. As a result, it becomes a resource which is shared among the symbols. In general, the amount of the introduced redundancy and thus the code rate can be chosen arbitrarily. In this study, we focus on constant redundancy per symbol and use code symbols represented by $n=8$~bits.

After processing all the symbols, the final encoder state should be communicated to the decoder. This state is required by the backward correction algorithm. When only forward correction is used, this step can be omitted at the cost of potential corruption of the last portion of the symbols.

The operation of the encoder is summarized in Algorithm~\ref{algorithm:encoder}. The details of the state transition procedure will be presented in Section~\ref{sec:state-transition}.

\begin{algorithm}[t]
\footnotesize{
\caption{Encoding procedure}
\label{algorithm:encoder}
\begin{algorithmic}
\REQUIRE $k,n,R := n-k$
\REQUIRE $\mathbf{x} := x_1, x_2, \ldots, x_l$ // Data stream $n-$bit symbols with $R$ bits zeroed

\STATE $s_0 \leftarrow $ Initialize system state
\STATE $\rmsk \leftarrow 2^{R}-1$ // Redundancy extraction mask

\FOR{$i = 1 \to M$}
  \STATE $s_i \leftarrow \mbox{updateState}(s_{i-1},x_i)$
  \STATE $y_i \leftarrow x_i + (s_i~\&~\rmsk)$
\ENDFOR

\STATE Transmit $\mathbf{y} := y_1, y_2, \ldots, y_M$ // Encoded stream of $n-$bit symbols
\end{algorithmic}
}
\end{algorithm}

\subsection{Forward Correction Procedure}

Principally, the decoder repeats the symbol processing procedure performed by the encoder. Upon reception of a new symbol, it performs a symbol-dependent state transition. From the posterior system state, it compares the resulting redundancy bits with the received ones. As a result, for error-free transmission, the decoding process is practically cost-free and takes the same amount of time as the encoding.

Assuming $\RR$ redundancy bits per symbol have been used, the probability of detecting an error is $p_d = 1 - 2^{-\RR}$ per node. When an error is successfully detected, the corresponding node of the correction tree is not analyzed any further. The decoder then considers other correction paths, starting from the most probable ones. We will later show that for a given $p_d$ and channel error rate below the Shannon limit, the decoder will be able to prune correction tree branches faster than they are created.

We refer to the posterior state which passed the redundancy check as \emph{allowed}, and the failing one as \emph{forbidden}. When a forbidden state is reached, the decoder needs to go back and apply a new correction to the current or one of previous symbols. Intuitively, if the current symbol is being considered for the first time in the current correction path, the most likely error pattern is a single bit error in the current symbol. In general, the decoder will select the next correction path for consideration by choosing a correction tree node with the greatest weight. The weights are calculated cumulatively from the beginning of the correction process and decremented each time a next bit is considered as corrupted.

A proper sequence of traversing the correction tree branches is ensured by using a heap for storing the tree nodes. The heap is ordered by the weights of the nodes. For the sake of the decoding performance, the decoding algorithm considers only feasible corrections, i.e., the ones that lead to the allowed states. This can be achieved by using a properly designed state transition procedure and dedicated state transition tables. Hence, it is possible to significantly reduce the use of the heap, which is the most time-consuming operation in the process. The forward error correction procedure is presented in a simplified way in Algorithm~\ref{algorithm:forward-correction}. The complete algorithm is included as a source code along with the paper~\cite{correction-trees-sources}.

\begin{algorithm}[t]
\footnotesize{
\caption{Operation of the decoding process with forward correction (simplified)}
\label{algorithm:forward-correction}
\begin{algorithmic}
\REQUIRE $k,n,R := n-k$
\REQUIRE $s_f$ // Final system state
\REQUIRE $\mathbf{y} = y_1, y_2, \ldots, y_l$ // Data stream $n-$bit symbols

\STATE Initialize correction lookup table
\STATE $s \leftarrow $ Initial system state
\STATE $cn \leftarrow null$ // The current node
\STATE $pn \leftarrow \mbox{before the first symbol}$ // Parent node

\REPEAT
\REPEAT
\STATE // Consider $pn.e$-th correction of parent node
\STATE Set $cn$ as $np.e$-th child of $pn$
\STATE $cn.e \leftarrow 0$  // Set it to null correction
\STATE $cn.s \leftarrow s$ // Remember current system state
\STATE Remember $cn$ on a list of visited nodes
\STATE $s,c \leftarrow \mbox{updateState}(s,y_{cn.d})$ // System state transition
\STATE $pn \leftarrow cn$ // Set current node as parent for the next cycle
\STATE // Finish if forbidden state is detected or there are no more symbols
\UNTIL{$c = false~\OR~cn.d > \mbox{last symbol}$}
\STATE // Set to next correction pattern passing verification
\STATE $cn.e \leftarrow \mbox{from pre-initialized lookup table}$
\STATE $\mbox{pushHeap}(cn)$ // Push the item on the heap
\STATE $pn \leftarrow \mbox{popHeap}()$ // Get new node to consider
\IF{$pn$ is null correction child}
\STATE // Create first  sibling of $pn$ and add it to heap
\STATE $cn \leftarrow pn.p$ // Set $cn$ as parent of $pn$
\STATE $cn.e \leftarrow \mbox{from pre-initialized lookup table}$
\STATE $\mbox{pushHeap}(cn)$ // Push the item on the heap
\ENDIF
\STATE Increase the number of error for considered node...
\STATE ...in the list of visited nodes and push it to heap
\STATE $s \leftarrow \mbox{updateState}(pn.s,y_{pn.d} \oplus pn.e)$ // System state transition
\STATE // Finish if after processing the last symbol the system state is correct
\UNTIL{$cn.d > \mbox{last symbol}~\AND~s = s_f$}
\end{algorithmic}
}
\end{algorithm}

\begin{figure*}[t]
    \centering
        \usetikzlibrary{calc,arrows,decorations.pathreplacing,decorations.pathmorphing,backgrounds,positioning,fit,patterns,shapes}

\tikzset{
    %Define standard arrow tip
    >=stealth',
    %Define style for boxes
    symbol/.style={
           circle,
           draw=black,
           fill=white,
           text centered},
    cor/.style={
          font = \footnotesize,
          anchor=south west,
          pos=0.3,
	  },
    % Define line style
    line/.style = {draw, ->},
    narline/.style = {text width=2cm, font = \small}
}

\begin{tikzpicture}[scale=0.65,transform shape]

\draw[help lines] (-1,5) -- (21,5) -- (21,-5) -- (-1,-5) -- (-1,5);
\draw[help lines,dotted] (-1,-5) grid[xstep=2,ystep=0.5] (21,5);
\draw[help lines] (-1,2.5) -- (21,2.5);

\draw[decorate,decoration={brace,raise=10,amplitude=5}] (-1,-5) -- (-1,5);
\node[rotate=90] at (-2.75,0) {All possible system states : $2^{64}$};

\draw[decorate,decoration={brace,raise=10,amplitude=5}] (21,5) -- (21,2.5);
\draw[decorate,decoration={brace,raise=10,amplitude=5}] (21,2.5) -- (21,-5);
\node[rotate=0,anchor=west,text width=3cm] at (22.5,-1.5) {Allowed system states, invalid with $1-p_d$};
\node[rotate=0,anchor=west,text width=3cm] at (22.5,3.5) {forbidden states, detected with $p_d$};

\draw[line] (5,5.5) -- (5,5); 
\draw[line] (11,5.5) -- (11,5); 
\draw[line] (19,5.5) -- (19,5); 
\node[anchor=south] at (5,5.5) {Error $c_1$};
\node[anchor=south] at (11,5.5) {Error $c_2$};
\node[anchor=south] at (19,5.5) {Error $c_1$};

\node[symbol] (s1) at (0,0) {$1$};

\node[symbol] (s2) at (2,1.5) {$2$};
\path[line] (s1) -- node[cor,anchor=south east,pos=0.7] {$s_1|c_0$} (s2);

\node[symbol] (s3) at (4,-3) {$3$};
\path[line] (s2) -- node[cor] {$s_2|c_0$} (s3);

\node[symbol] (s4) at (6,4) {$4$};
\path[line,dashed] (s3) -- node[cor] {$s_3|c_0$} (s4);

\node[symbol] (s5) at (6,-1) {$5$};
\path[line] (s3) -- node[cor,anchor=north west,pos=0.4] {$s_3|c_1$} (s5);

\node[symbol] (s6) at (8,-3.4) {$6$};
\path[line] (s5) -- node[cor,anchor=north east,pos=0.5] {$s_4|c_0$} (s6);

\node[symbol] (s7) at (10,1) {$7$};
\path[line] (s6) -- node[cor,anchor=south east,pos=0.7] {$s_5|c_0$} (s7);

\node[symbol] (s8) at (12,3) {$8$};
\path[line,dashed] (s7) -- node[cor,anchor=south east,pos=0.5] {$s_6|c_0$} (s8);

\node[symbol] (s9) at (12,0) {$9$};
\path[line,dashed] (s7) -- node[cor] {$s_6|c_1$} (s9);

\node[symbol] (s10) at (14,4) {$10$};
\path[line,dashed] (s9) -- node[cor,anchor=north west] {$s_7|c_0$} (s10);

\node[symbol] (s11) at (14,-2) {$11$};
\path[line,dashed] (s9) -- node[cor] {$s_7|c_1$} (s11);

\node[symbol] (s12) at (16,3.5) {$12$};
\path[line,dashed] (s11) -- node[cor] {$s_8|c_0$} (s12);

\node[symbol] (s13) at (12,-3) {$13$};
\path[line] (s7) -- node[cor,pos=0.7] {$s_6|c_2$} (s13);

\node[symbol] (s14) at (14,-4.25) {$14$};
\path[line] (s13) -- node[cor,pos=0.5] {$s_7|c_0$} (s14);

\node[symbol] (s15) at (16,-2.25) {$15$};
\path[line] (s14) -- node[cor,pos=0.6,anchor=south east] {$s_8|c_0$} (s15);

\node[symbol] (s16) at (18,-1) {$16$};
\path[line] (s15) -- node[cor,pos=0.7,anchor=south east] {$s_9|c_0$} (s16);

\node[symbol] (s17) at (20,-4) {$17$};
\path[line,dashed] (s16) -- node[cor,anchor=south west,pos=0.4] {$s_{10}|c_0$} (s17);

\node[symbol] (s18) at (20,1.5) {$18$};
\path[line] (s16) -- node[cor,anchor=north west,pos=0.4] {$s_{10}|c_1$} (s18);

% 
% \node[symbol,right= of s3] (s4) {$s_4$};
% \path[line] (s3) -- node[cor] {c=0} (s4);
% 
% \node[symbol,below= of s4] (s4c1) {$s_4$};
% \path[line] (s3) to [in=150,out=-60] node[cor,anchor=south west] {c=1} (s4c1);
% 
% \node[symbol,below= of s4c1] (s4c2) {$s_4$};
% \path[line] (s3) to [in=150,out=-80] node[cor,anchor=south west,pos=0.7] {c=2} (s4c2);
% 
% \node[symbol,right= of s4c2] (s5) {$s_5$};
% \path[line] (s4c2) -- node[cor] {c=0} (s5);
% 
% \node[symbol,right= of s5] (s6) {$s_6$};
% \path[line] (s5) -- node[cor] {c=0} (s6);
% 
% \node[symbol,right= of s6] (s7) {$s_7$};
% \path[line] (s6) -- node[cor] {c=0} (s7);
% 
% \node[symbol,right= of s4c1] (s5v2) {$s_5$};
% \path[line] (s4c1) -- node[cor] {c=0} (s5v2);

\end{tikzpicture}
        \caption{Illustration of the forward error correction process and the corresponding system state transitions. Solid lines represent the proper correction path.}
        \label{fig:pathtrace}
\end{figure*}
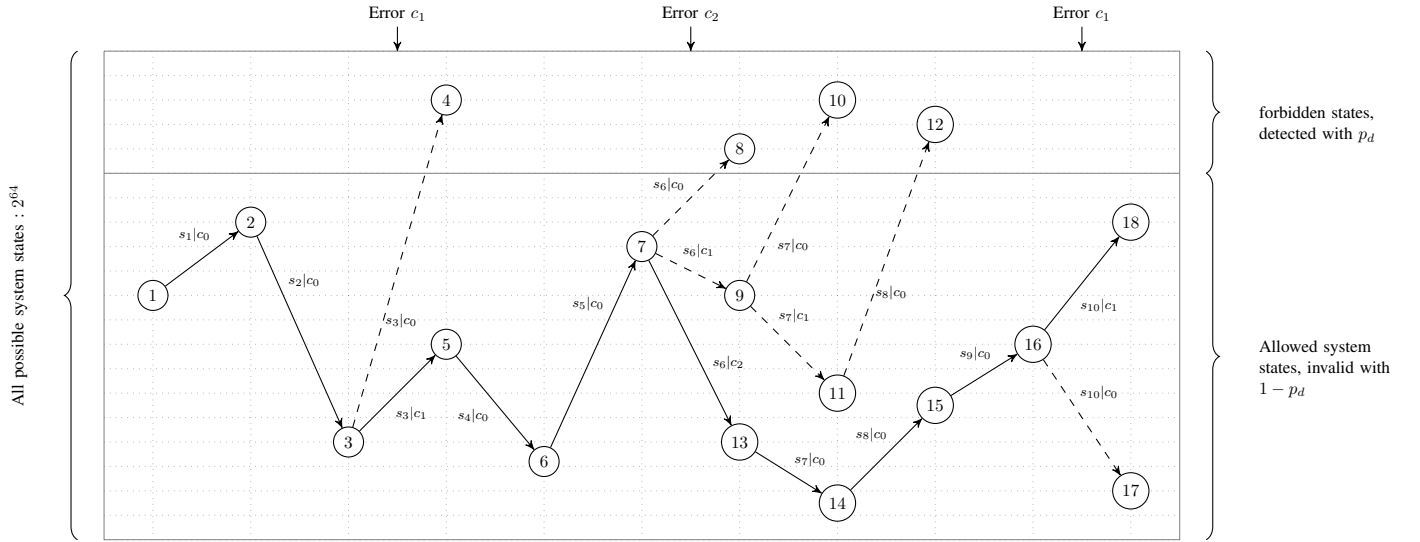

The considered correction method is illustrated schematically in Fig.~\ref{fig:pathtrace}. The numbers of the nodes represent the sequence in which the decoder traverses the correction tree. The labels of the state transitions denote the symbol and the currently considered correction of that symbol, i.e., $c_0$ denotes a null correction vector, $c_1$ the most probable correction vector, $c_2$ the second probable correction vector, etc. In the presented example the $s_3$, $s_6$ and $s_{10}$ symbols are corrupted. For $s_3$, the decoder immediately notices a forbidden system state and the most probable correction turns out to be the proper one. For $s_6$, the decoder detects a forbidden state at first, but then fails to correct it. After considering a number of wrong corrections, the most probable turns out to be the second correction of $s_6$. For the last symbol, $s_{10}$, the decoder does not rely on its redundancy information and uses the knowledge of the final system state to detect wrong correction vectors with probability close to $1$.

\subsection{Weighting the Correction Tree Nodes}
\label{sec:weights}

In this study, we consider the transmission channel to be a binary symmetric channel, i.e., for each of the transmitted bits, there is a constant probability ($\pb > 0$) that the bit will be received as corrupted. The number of possible errors in a $N$ bit sequence can be asymptotically estimated as:

\be {N \choose \pb N}\approx 2^{Nh(\pb)} \ee

If we were to build a tree of all typical corrections for $j$ successive symbols ($j\nn$ bits), it should contain asymptotically $2^{j\nn h(\pb)}$ leaves. If the probability of detecting an error is indeed $p_d$, on average only $(1-p_d)^j$ of incorrect of them should survive the associated $j$ redundancy tests. Then, the expected number of survivors can be asymptotically estimated as:

% \begin{eqnarray}
%  2^{j \nn h(\pb)} (1-p_d)^j & = & \left(2^{\nn h(\pb)+\lg(\tilde{p_d})}\right)^j \\
%  & = & \left(2^{\nn h(\pb)-\RR}\right)^j \\
% \end{eqnarray}

\begin{equation}
 2^{j \nn h(\pb)} (1-p_d)^j = \left(2^{\nn h(\pb)+\lg(\tilde{p_d})}\right)^j = \left(2^{\nn h(\pb)-\RR}\right)^j \\
\end{equation}

Thus, codes with rates $R > R_0 := \nn h(\pb)$ or equivalently
\begin{equation}
 p_d > p_d^0 := 1 - 2^{-R_0}
\end{equation}
\noindent are correctable in the sense that the correction tree no longer grows exponentially. This is in fact the Shannon theoretical limit for the binary symmetric channel. Under this limit, a fixed large system state space is sufficient to guarantee that there are practically no state collisions during the correction process.

Selection of the most likely correction path is based on Bayesian analysis:

$$\Pr(E|O)\propto \Pr(O|E)\Pr(E)$$

\noindent where observation ($O$) in our case is the constructed tree. The explanation ($E$) we are looking for is the proper correction (leaf of the tree). We will denote such correction of $j$ symbols ($J=\nn j$ bits) using the bit vector $(E_i)_{i=1}^J : E_i=1$ if and only if the $i^{th}$ bit is changed. The probability of the correction $E$ can be directly calculated from the definition of the BSC as:

$$\Pr(E)=\pb^{\#\{i:E_i=1\}}(1-\pb)^{\#\{i:E_i=0\}}$$

If the given correction (explanation) is proper, the tree nodes which are not on the correction path correspond to wrong corrections. The probability of obtaining the current situation when $E$ is indeed the proper path is
$$\Pr(O|E)=p_d^{f}(1-p_d)^{a} $$

\noindent where $f$ is the number of forbidden nodes outside the currently considered path and $a$ is the number of allowed among them. By dividing this expression by an analogous one for all nodes of the tree, there remains only the dependency on the number of nodes in the currently considered correction path, i.e., $(1-p_d)^{-j}$. Finally, by taking the logarithm of $Pr(O|E) \cdotp Pr(E)$, we see that for a given correction tree, the next most probable correction path to consider corresponds to a node which maximizes:

\be \#\{i:E_i=1\}\lg(\pb)+\#\{i:E_i=0\}\lg (\tilde{\pb})-j\lg (\tilde{p_d}) \label{eq:weight}\ee

\noindent The first two terms favor corrections with smaller number of corrected bits along the path. The last term favors longer paths. \eqref{eq:weight} represents the weight of a correction tree node and can be seen as a form of the Fano metric \cite{fano}. It is expressed as a logarithm of probability, therefore a difference of the weight of 1 denotes a node twice as probable than the other.

Although it is not the focus of this study, it is worth mentioning that this method can also be adapted to other types of communication channels. An erasure channel can be modeled by using $\pb=1/2$ for the erased bits. The bit error probability can also be calculated in a per-bit manner to take into account soft-decisions of digital transmission demodulators. By increasing the set of possible symbol corrections, such an approach would also be capable of handling synchronization errors, such as bit deletion or duplication, which are difficult to deal with in other error correction methods.

In practice, \eqref{eq:weight} is calculated in a cumulative manner, i.e., the weight is calculated for the currently considered symbol only and it is added to the weight of its parent.

\subsection{System State Transitions}
\label{sec:state-transition}

This section describes the designed symbol-dependent system state transition procedure. The main design objective was to allow for low cost computation and to ensure good distance and statistical properties for correction in both directions.

A single transmitted symbol contains $\nn$ bits of information, including $\kk$ payload bits and $\RR$ redundancy bits. In this study, we consider only natural $\RR$; however, this derivation can be made more general, for example the redundancy can be added while entropy coding by introducing forbidden symbols of probability $p_d$ to the original alphabet \cite{ans}.

Upon reception of a symbol, the state of the system should be changed according to the received payload bits. For this purpose we perform a bit-wise exclusive disjunction of the current system state with a \emph{transition vector} for the given payload. These transition vectors are constant throughout the process, and they are stored in a lookup table for the sake of operation performance. In the final step, the current state is shifted cyclically by $\RR$ bits.

The system state update algorithm for forward correction is shown in Algorithm~\ref{algorithm:update-forward} and for backward correction in Algorithm~\ref{algorithm:update-backward}.

\begin{algorithm}[t]
\footnotesize{
\caption{Forward state transition procedure}
\label{algorithm:update-forward}
\begin{algorithmic}
\REQUIRE $s_{i-1}$ // Previous state
\REQUIRE $y_i$ // New symbol
\REQUIRE $\mathbf{t} = t_1, t_2, \ldots, t_{2^k}$ // State transition table

\STATE $\rmsk \leftarrow 2^{R}-1$ // Redundancy extraction mask
\STATE $\ttemp \leftarrow t_{y_i >> R}$ // Temporary variable
\STATE $c \leftarrow ((s_{i-1}\oplus y_i\oplus \ttemp)~\&~\rmsk) = 0$ // Redundancy verification result
\STATE $s_i \leftarrow ((s_{i-1}\oplus \ttemp)>>R)+(s_{i-1} << (64-R)) $ // Apply state transition

\end{algorithmic}
}
\end{algorithm}

\begin{algorithm}[t]
\footnotesize{
\caption{Backward state transition procedure}
\label{algorithm:update-backward}
\begin{algorithmic}
\REQUIRE $s_{i-1}$ // Previous state
\REQUIRE $y_i$ // New symbol
\REQUIRE $\mathbf{t} = t_1, t_2, \ldots, t_{2^k}$ // State transition table

\STATE $\rmsk \leftarrow 2^{R}-1$ // Redundancy extraction mask
\STATE $\ttemp \leftarrow t_{y_i >> R}$ // Temporary variable
\STATE $s_i \leftarrow ((s_{i-1}\oplus \ttemp)<<R)+(s_{i-1} >> (64-R)) $ // Apply state transition
\STATE $c \leftarrow ((s_{i}\oplus y_i\oplus \ttemp)~\&~\rmsk) = 0$ // Redundancy verification result

\end{algorithmic}
}
\end{algorithm}

Strictly pseudo-random initialization of the transition vectors is possible; however, a conscious choice enables significant improvements in the number of the decoding steps required. Firstly, since $\RR$ least significant bits of the system state are directly used as the redundancy, it is of crucial importance that they allow for immediate detection of a forbidden system state. For this purpose, we evaluated all possible symbol correction patterns and selected $\RR$ redundancy bits, so that the probability of missing these errors would be minimal. For example, for a code with $\RR = 1$ (code rate 7/8,) it is intuitively clear that the single bit of redundancy should be an exclusive disjunction on all of the payload bits. The corresponding matrix representation is:

$$\left(\begin{array}{ccccccc}  1 & 1 & 1 & 1 & 1 & 1 & 1  \\ \end{array}  \right) $$

It immediately detects single bit errors, although it fails to detect double errors. For $\RR=4$ (1/2 rate) the best matrix is:

$$\left(\begin{array}{cccc} 0 & 1 & 1 & 1 \\  1 & 0 & 1 & 1 \\  1 & 1 & 0 & 1 \\   1 & 1 & 1 & 0 \\ \end{array}  \right)$$

It detects all damages up to the triple bit and 56 of 70 quadruples. For $\RR=6$ (1/4 rate) the matrix

$$\left(\begin{array}{cccccc} 1 & 1 & 1 & 1 & 0 & 0 \\ 1 & 1 & 0 & 0 & 1 & 1 \\ \end{array} \right)^T $$

\noindent immediately detects all up to the quadruple and 54 of 56 quintuple bit damages.

In order to maximize the amount of information carried by the redundancy bits, the number of payload bit sequences that correspond to 0 should be equal to the ones corresponding to 1 on each position of the remaining $64-\RR$ bits of the transition vectors. Moreover, the remaining bits should be made as independent as possible. For this purpose, we divide the remaining $64-\RR$ bits into $\kk$ length segments, and initialize each segment with independent pseudo-random permutations of the possible $2^{\kk}$ payload bit sequences. This approach is a significant improvement on the pseudo-random choice of the whole transition vector. However, we believe that it could be optimized even further.

Together with the selection of the initial system state, utilization of pseudo-random numbers in the state transition process allows us to require the knowledge of the cryptographic key to utilize the introduced redundancy.

\paragraph{Optimization of the Considered State Space}

In order to optimize the operation of the proposed decoder and severely reduce the amount of the considered correction patterns, we automatically discard the patterns which lead to incorrect system states. For each considered symbol, there exists only a small subset (approx. $2^{-R}$) of correction patterns which lead to colliding redundancy bits of the posterior system state. For the sake of optimal performance, the decoder should consider only these corrections. The decoder pre-initializes a lookup table of such correction patterns and orders them by their weights. Such an approach successfully reduces the number of operations on the heap and significantly improves the correction performance.

In the forward decoding step the condition for the redundancy verification result $c$ can be rewritten as:

\begin{equation}
 (y_i \oplus s_{i-1})~\&~\rmsk = t_{y_i>>R}~\&~\rmsk  \label{cond}
\end{equation}
% \be \verb"(symbol ^ state) & \rmskask == transition[symbol >> R] & \rmskask" \label{cond}\ee

Let us introduce a \emph{modified symbol} $z_i$:

\begin{equation}
 z_i = y_i \oplus (s_{i-1}~\&~\rmsk)
\end{equation}
% \be \verb"modified_symbol = symbol ^ (state & \rmskask)"\ee

Applying the error mask to the modified symbol corresponds to applying it to the original symbol. However, it eliminates the dependency on the state from the original condition (\ref{cond}):

\begin{equation}
 z_i~\&~\rmsk = t_{z_i>>R}~\&~\rmsk
\end{equation}
% \be \verb"modified_symbol & \rmskask == transition[symbol >> R] & \rmskask" \ee

This representation is more convenient to use in the lookup process. For example, we use it to generate the lookup table of the allowed error patterns for each possible $z_i\in[0,2^{{\nn}}-1]$. Cyclic shift was modified to make it possible to use the same redundancy bits for both forward and backward correction. During the latter:

\begin{equation}
 z_i = y_i \oplus (s_{i-1} >> 64 - R)
\end{equation}
% \be \verb"modified_symbol = symbol ^ (state >> (state_size - R))"\ee

\subsection{Bidirectional Correction}

Simultaneous forward and backward correction enables significantly better correction performance, as at least two CECs are required to cripple the correction process. When using bidirectional correction, the decoder builds two correction trees which are expanded independently as long as the symbol positions do not overlap. When this happens, the decoder looks for the identical system states for overlapping symbols. If this is successful, the correction paths are merged and the decoder yields the correct data stream.

In our implementation, we use a single array for storing the structure of both trees, and the decoder performs cyclically one correction step per direction. For each of the overlapping symbols, we create a binary search tree in order to obtain a logarithmic operation time. This way, the decoder can simply proceed with optimal expansion of both correction directions. An alternative approach would be to establish a barrier at the first overlapping symbol and make the decoder focus on finding the matching state there. However, the meeting point is usually placed asymmetrically in a CEC region and the resources are wasted in the direction which has already reached the proper correction. Then, the meeting provides almost no help with the difficult correction from the second direction. It is possible to shift the meeting point, although the available information is insufficient to do it optimally.

\subsection{Final State}
\label{sec:final-state}

The proposed correction method requires to know the final system state in order to achieve the best correction performance. Without the final state, the decoder is not capable of backward correction. There is still the possibility of forward correction, which has a slightly lower efficiency when used alone.

The final state should be transmitted to the decoder in a reliable way. In this study, we do not address this issue. However, the decoder could easily be extended to support a small number of corrupted bits in the final system state. Instead of starting the backward correction from a single state, the heap of possible corrections should be pre-initialized with all reasonably probable correction patterns. Such a mechanism operates in the same way as ordinary correction of the transmitted symbols. For efficient operation, the weights of these alternative final states should be adjusted and calculated with some estimated bit error probability $\pb'<<\pb$. The probability that a given bit should not be changed is $(1-\pb')/\pb'$ times larger than the opposite hypothesis; therefore, for each changed bit, the initial weight of such modified states should be lower by $\lg((1-\pb')/\pb')$.

For the sake of reliable transmission of this state information, an arbitrary error correction mechanism can be used. However, since state-of-the-art correction algorithms are not efficient when working on small data blocks, we propose to aggregate the final states in large blocks and use the proposed correction mechanism. Alternative solution is placing the final states in the beginning of the succeeding frame (before encoding) - the vicinity to the initial state makes this region much more damage resistant. Such sequence of frames has to be decoded in backward order.

This additional communication overhead causes a slight drop in the effective code rate of the proposed method. For example if we assume that for rate 1/2 the state is protected using the same rate (on average $8\cdot 2 =16$ bytes), for 1024 byte frames the real rate is $512/(1024+16)\approx 0.4923$.

\section{Theoretical analysis}
\label{sec:theory}

In this section, we analyze the asymptotic statistical behavior of the proposed correction method. We quantitatively evaluate error concentrations using weight drops, as well as their probability distributions and average correction costs for BSC.

Both the decrease of the probability of error concentrations and the growth of the correction cost are usually exponential. Our analysis yields a theoretical rate of the exponents, allowing us to find the probability distribution of correction failure for a chosen step limit for both uni- and bidirectional correction. Our analysis is verified by experimental evaluation, described in detail in Section~\ref{sec:evaluation}.

For better intuition with respect to the described behavior, it might be useful to experiment with an interactive simulator of the proposed correction tree approach \cite{wolfram}.

\subsection{Single-direction Sequential Correction}

Asymptotically, the average weight (\ref{eq:weight}) per node for a proper correction path is:

$$\nn\pb \lg(\pb) + \nn\tilde{\pb}\lg(\tilde{\pb})-\lg(\tilde{p}_d)=-\nn h(\pb)-\lg(\tilde{p}_d)$$

Therefore, the condition of using a code rate below the Shannon limit ($1-h(\pb)>\frac{\kk}{\nn}=\frac{\nn+\lg(\tilde{p}_d)}{\nn}$) is equivalent to that the weight of the proper correction path is statistically growing. Locally, however, it can decrease. In such situations, before finding the proper correction path again, the decoder needs to expand some of the wrong correction sub-tree.

\begin{figure*}[t]
    \centering
        \includegraphics[width=5in]{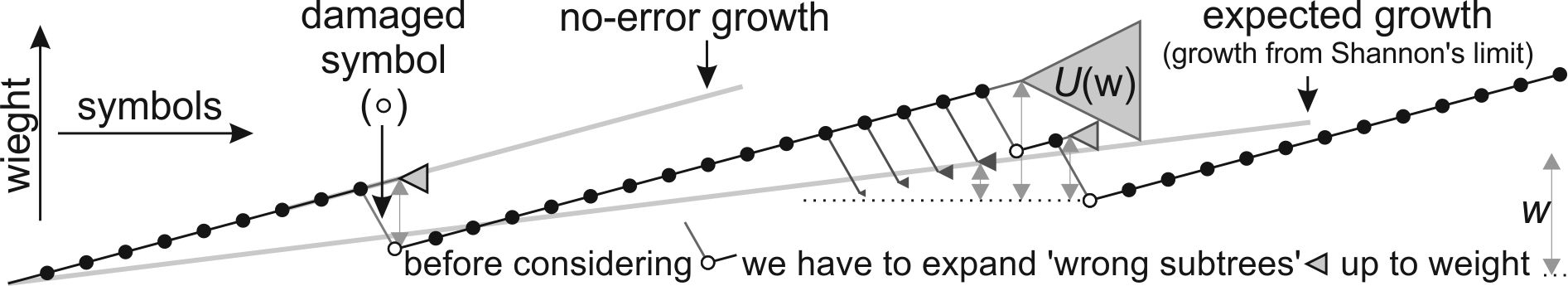}
        \caption{Schematic diagram of the proper correction we aim to find. After an error, we expand the wrong corrections until reaching the weight of the proper one. We search for expected size and probability of occurrence for these wrong correction subtrees.}
        \label{an}
\end{figure*}

The situation is outlined in Fig. \ref{an}. For the proper correction path, let us define the \emph{weight drop} $w$ for a given node as its weight minus the minimum from the weights of all future nodes, i.e. weight drop equals 0 if the weight does not become smaller. We will not use the node with this minimum weight until we expand trees of wrong corrections up to this weight drop. This means that it quantitatively describes error concentrations: not only the number of damaged bits, but also how densely they are distributed. The expected number of such wrong corrections that need to be considered will grow exponentially with $w$, but the probability of large $w$ will drop exponentially with it.

Firstly, let us consider the weight drop probability distribution. Specifically, for an infinite correct path, let us define a function $V(w)$ as the probability that the weight of further nodes (including this node) will drop by at most $w$. For $w<0,\ V(w)=0$. $V(0)$ corresponds to a situation in which the weight increases and it should be positive. This function is not continuous, but surprisingly it asymptotically tends to a continuous function.

Before analyzing a general case, let us focus for a moment on a simplified one. Let us assume that all symbols have exactly 1 bit ($\nn=1$) of information. Observe that we can write $V$ using its values for the previous position, getting a functional equation (\ref{veqv}) for BSC.

\begin{figure*}[!t]
% ensure that we have normalsize text
\normalsize
% Store the current equation number.
\setcounter{MYtempeqncnt}{\value{equation}}
% Set the equation number to one less than the one
% desired for the first equation here.
% The value here will have to changed if equations
% are added or removed prior to the place these
% equations are referenced in the main text.
\setcounter{equation}{8}
$$ V(w):=\textrm{probability that the weight on the correct path will drop by at most }w  $$
\be V(w)=\left\{\begin{array}{ll} \pb\, V(w+\lg(\pb)-\lg(\tilde{p}_d))+\tilde{\pb}\, V(w+\lg(\tilde{\pb})-\lg(\tilde{p}_d))\quad & \textrm{for }w\geq 0\\
                 0 & \textrm{for }w<0 \end{array}\right. \label{veqv}\ee
$$U(w):=\textrm{expected number of processed nodes of wrong correction from node of weight drop }w$$
\be U(w)=\left\{\begin{array}{ll} 1+\tilde{p}_d\left( U(w+\lg(\pb)-\lg(\tilde{p}_d))+ U(w+\lg(\tilde{\pb})-\lg(\tilde{p}_d))\right)\quad & \textrm{for }w\geq 0\\
                 0 & \textrm{for }w<0 \end{array}\right. \label{vequ}\ee

% Restore the current equation number.
% \setcounter{equation}{\value{MYtempeqncnt}}
\setcounter{equation}{10}
% IEEE uses as a separator
\hrulefill
% The spacer can be tweaked to stop underfull vboxes.
\vspace*{4pt}
\end{figure*}

If there was no such boundary behavior in $w=0$, it would be a simple linear functional equation with a linear combination of exponents as a solution. Fortunately the functional equation is sufficient to find the asymptotic behavior.

We know that $\lim_{w\to\infty} V(w)=1$, so let us assume that asymptotically

\be 1-V(w)\propto 2^{vw}\label{subs1}\ee

\noindent for some $v<0$. Substituting it to (\ref{veqv}), we get:

$$2^{vw}=\pb\, 2^{v(w+\lg(\pb)-\lg(\tilde{p}_d))}+\tilde{\pb}\, 2^{v(w+\lg(\tilde{\pb})-\lg(\tilde{p}_d))}$$

\be \tilde{p}_d^v=\pb^{v+1} +\tilde{\pb}^{v+1} \label{equv}\ee

This equation always has a $v=0$ solution, but for $p_d>p_d^0$ there emerges a second, negative solution which is of interest here. It can be easily found numerically and our simulations show that we are asymptotically reaching this behavior.

We can now go to the general case. By writing (\ref{veqv}) analogously for a larger natural $\nn$, we get $2^{\nn}$ terms and after the substitution (\ref{subs1}), we can collapse them:

$$2^{vw}=2^{vw}\left(\pb\, 2^{v(\lg(\pb)-\frac{1}{\nn}\lg(\tilde{p}_d))}+\tilde{\pb}\, 2^{v(\lg(\tilde{\pb})-\frac{1}{\nn}\lg(\tilde{p}_d))}\right)^{\nn}$$
\be \tilde{p}_d^v=\left(\pb^{v+1} +\tilde{\pb}^{v+1}\right)^{\nn} \label{vvrel} \ee

Again we are interested in the $v<0$ solution. \\

Next we need to estimate the asymptotic behavior of the size of the wrong correction sub-trees for large weight drops. Define $U(w)$ as the expected number of nodes of such a sub-tree, which would be constructed from a node of weight drop $w$. Each node of such a sub-tree can be thought of as a root of a new sub-tree. If we expand it for the corresponding weight drops, we obtain the expected number of nodes. Again, $U(w)=0$ for $w<0$. For $w=0$ we process this node, i.e., $U(0)\geq 1$.

First, let us focus on one bit blocks ($\nn=1$) as previously. Connecting the node with its children we get the functional equation (\ref{vequ}). We expect that for some $u>0$ asymptotically

\be U(w)\propto 2^{uw} \label{eq:uwprop} \ee

By substituting (\ref{eq:uwprop}) to (\ref{vequ}) we obtain

$$2^{uw}=\tilde{p}_d\left(2^{u(w+\lg(\pb)-\lg(\tilde{p}_d))}+
2^{u(w+\lg(\tilde{\pb})-\lg(\tilde{p}_d))}\right)$$
\be \tilde{p}_d^{u-1}=\pb^u+\tilde{\pb}^u\label{urel}\ee

This equation is very similar to (\ref{equv}). For $p_d>p_d^0$ we once again obtain two solutions, with the greater one equal to 1. Analogously to the previous analysis, due to strong boundary conditions in $w=0$, we will be asymptotically reaching the smaller solution. This behavior is confirmed by our simulations. By comparing these two equations, we surprisingly obtain a simple correspondence between these two critical coefficients:

\be u=v+1 \qquad\qquad\qquad(v<0,\ u>0)\label{rel}\ee

This relationship is also satisfied in the general case ($\nn>1$). If $\pb$ does not match the channel's properties exactly, $U$ would remain the same, but $V$ would be slightly different.Therefore this relationship would be approximate only. \\

\begin{figure*}[t!]
 \centering
  % This file was created by matlab2tikz v0.0.5.
% Copyright (c) 2008--2010, Nico Schlömer <nico.schloemer@ua.ac.be>
% All rights reserved.
%
% The latest updates can be retrieved from
%  http://win.ua.ac.be/~nschloe/content/matlab2tikz/
% and
%  http://www.mathworks.com/matlabcentral/fileexchange/22022 .
% where you can also make suggestions and rate matlab2tikz.

\begin{tikzpicture}[font=\footnotesize]

% Axis at [0.13 0.11 0.78 0.79]
\begin{axis}[
scale only axis,
width=5in,
height=2.5in,
xmin=0, xmax=0.30,
ymin=1, ymax=8,
xlabel={$\pb$},
 ylabel style={yshift={-0.05in}},
 ylabel={$\frac{1}{rate}$},
axis on top,
extra description/.code = {
  \node at (0.1, 0.2) {$c = -1$};
  \node at (0.865, 0.5) {$c = 0$};
 },
xtick={0,0.05,0.1,0.15,0.2,0.25,0.3}
]

\addplot [
color=black,
dashed,
]
coordinates{
(0, 2)
(0.4, 2)
};

\addplot [
color=black,
dashed,
]
coordinates{
(0, 3)
(0.4, 3)
};

\addplot [
color=black,
dashed,
]
coordinates{
(0, 4)
(0.4, 4)
};

\addplot [
color=black,
dashed,
]
coordinates{
(0, 1.142857143)
(0.4, 1.142857143)
};

\addplot [
color=black,
solid,
]
coordinates{
(0, 1.)
(0.010001, 1.0879)
(0.020001,  1.16475)
(0.030001, 1.24131)
(0.040001, 1.31978)
(0.050001,  1.40135)
(0.060001, 1.48688)
(0.070001, 1.57711)
(0.080001,  1.67275)
(0.090001, 1.77454)
(0.100001, 1.88323)
(0.110001,  1.99968)
(0.120001, 2.12478)
(0.130001, 2.25959)
(0.140001,  2.40524)
(0.150001, 2.56307)
(0.160001, 2.73457)
(0.170001,  2.92147)
(0.180001, 3.12577)
(0.190001, 3.34979)
(0.200001,  3.59622)
(0.210001, 3.86824)
(0.220001, 4.16961)
(0.230001,  4.50477)
(0.240001, 4.87905)
(0.250001, 5.29885)
(0.260001,  5.77194)
(0.270001, 6.3078)
(0.280001, 6.91813)
(0.290001,  7.61745)
};

\addplot [
color=black,
solid,
]
coordinates{
(0, 1.)
(0,  1.1)
(0.0037634, 1.2)
(0.00757958, 1.3)
(0.0121391,  1.4)
(0.0171851, 1.5)
(0.0225363, 1.6)
(0.0280641,  1.7)
(0.0336765, 1.8)
(0.0393075, 1.9)
(0.0449101,  2.)
(0.0504508, 2.1)
(0.055906, 2.2)
(0.0612592,  2.3)
(0.0664997, 2.4)
(0.0716204, 2.5)
(0.0766173,  2.6)
(0.0814885, 2.7)
(0.0862339, 2.8)
(0.0908545,  2.9)
(0.095352, 3.)
(0.0997289, 3.1)
(0.103988, 3.2)
(0.108133,   3.3)
(0.112166, 3.4)
(0.116092, 3.5)
(0.119914,  3.6)
(0.123635, 3.7)
(0.127258, 3.8)
(0.130788, 3.9)
(0.134227,   4.)
(0.137579, 4.1)
(0.140847, 4.2)
(0.144034, 4.3)
(0.147143,   4.4)
(0.150176, 4.5)
(0.153137, 4.6)
(0.156027, 4.7)
(0.15885,   4.8)
(0.161608, 4.9)
(0.164303, 5.)
(0.166938, 5.1)
(0.169513,   5.2)
(0.172032, 5.3)
(0.174497, 5.4)
(0.176908,  5.5)
(0.179269, 5.6)
(0.18158, 5.7)
(0.183844, 5.8)
(0.186061,  5.9)
(0.188234, 6.)
(0.190363, 6.1)
(0.19245, 6.2)
(0.194497,  6.3)
(0.196505, 6.4)
(0.198474, 6.5)
(0.200406, 6.6)
(0.202303,   6.7)
(0.204164, 6.8)
(0.205992, 6.9)
(0.207787, 7.)
(0.20955,  7.1)
(0.211283, 7.2)
(0.212985, 7.3)
(0.214658, 7.4)
(0.216302,   7.5)
(0.217919, 7.6)
(0.219509, 7.7)
(0.221072, 7.8)
(0.22261,   7.9)
(0.224124, 8.)
};

\addplot [
color=black,
solid,
]
coordinates{
(0, 1.)
(0.00363778,  1.1)
(0.00980341, 1.2)
(0.0170088, 1.3)
(0.0246774,  1.4)
(0.0325032, 1.5)
(0.0403105, 1.6)
(0.0479957,  1.7)
(0.0554976, 1.8)
(0.0627817, 1.9)
(0.0698299,  2.)
(0.0766347, 2.1)
(0.0831953, 2.2)
(0.0895156,  2.3)
(0.0956019, 2.4)
(0.101462, 2.5)
(0.107105,  2.6)
(0.112541, 2.7)
(0.117779, 2.8)
(0.122828, 2.9)
(0.127697,   3.)
(0.132396, 3.1)
(0.136933, 3.2)
(0.141316, 3.3)
(0.145552,   3.4)
(0.149649, 3.5)
(0.153614, 3.6)
(0.157453,  3.7)
(0.161172, 3.8)
(0.164777, 3.9)
(0.168274, 4.)
(0.171667,  4.1)
(0.174961, 4.2)
(0.178161, 4.3)
(0.18127, 4.4)
(0.184294,  4.5)
(0.187235, 4.6)
(0.190097, 4.7)
(0.192884, 4.8)
(0.195599,   4.9)
(0.198244, 5.)
(0.200823, 5.1)
(0.203338, 5.2)
(0.205792,   5.3)
(0.208187, 5.4)
(0.210525, 5.5)
(0.212809, 5.6)
(0.21504,   5.7)
(0.217221, 5.8)
(0.219353, 5.9)
(0.221439, 6.)
(0.223479,   6.1)
(0.225475, 6.2)
(0.227429, 6.3)
(0.229343,  6.4)
(0.231217, 6.5)
(0.233054, 6.6)
(0.234853, 6.7)
(0.236617,   6.8)
(0.238347, 6.9)
(0.240043, 7.)
(0.241707, 7.1)
(0.24334,  7.2)
(0.244942, 7.3)
(0.246515, 7.4)
(0.248059, 7.5)
(0.249576,   7.6)
(0.251066, 7.7)
(0.25253, 7.8)
(0.253968, 7.9)
(0.255382,   8.)
};

\addplot [
color=black,
solid,
]
coordinates{
(0, 1.)
(0.0065645, 1.1)
(0.0156218,  1.2)
(0.0253197, 1.3)
(0.0350942, 1.4)
(0.0446903,  1.5)
(0.0539851, 1.6)
(0.0629216, 1.7)
(0.071478,  1.8)
(0.0796522, 1.9)
(0.0874525, 2.)
(0.0948935,  2.1)
(0.101992, 2.2)
(0.108768, 2.3)
(0.115238, 2.4)
(0.121422,   2.5)
(0.127337, 2.6)
(0.132999, 2.7)
(0.138425,  2.8)
(0.143628, 2.9)
(0.148623, 3.)
(0.153422, 3.1)
(0.158037,  3.2)
(0.162478, 3.3)
(0.166756, 3.4)
(0.17088, 3.5)
(0.174858,  3.6)
(0.1787, 3.7)
(0.182411, 3.8)
(0.185999, 3.9)
(0.189471,  4.)
(0.192833, 4.1)
(0.196089, 4.2)
(0.199246, 4.3)
(0.202308,  4.4)
(0.205279, 4.5)
(0.208165, 4.6)
(0.210969, 4.7)
(0.213694,   4.8)
(0.216345, 4.9)
(0.218924, 5.)
(0.221435, 5.1)
(0.223881,   5.2)
(0.226264, 5.3)
(0.228586, 5.4)
(0.230852,  5.5)
(0.233062, 5.6)
(0.235218, 5.7)
(0.237324, 5.8)
(0.239381,   5.9)
(0.24139, 6.)
(0.243355, 6.1)
(0.245275, 6.2)
(0.247153,  6.3)
(0.248991, 6.4)
(0.250789, 6.5)
(0.252549, 6.6)
(0.254273,   6.7)
(0.255961, 6.8)
(0.257616, 6.9)
(0.259237, 7.)
(0.260826,   7.1)
(0.262385, 7.2)
(0.263913, 7.3)
(0.265413,  7.4)
(0.266884, 7.5)
(0.268329, 7.6)
(0.269747, 7.7)
(0.271139,   7.8)
(0.272506, 7.9)
(0.273849, 8.)
};

\addplot [
color=black,
solid,
]
coordinates{
(0,  1.)
(0.00873955, 1.1)
(0.019643, 1.2)
(0.0308389,  1.3)
(0.0418287, 1.4)
(0.0524145, 1.5)
(0.0625189,  1.6)
(0.0721203, 1.7)
(0.0812246, 1.8)
(0.0898512,  1.9)
(0.0980257, 2.)
(0.105776, 2.1)
(0.113131, 2.2)
(0.120117,   2.3)
(0.12676, 2.4)
(0.133085, 2.5)
(0.139114, 2.6)
(0.144867,   2.7)
(0.150364, 2.8)
(0.155623, 2.9)
(0.160658, 3.)
(0.165484,   3.1)
(0.170116, 3.2)
(0.174565, 3.3)
(0.178843,  3.4)
(0.182959, 3.5)
(0.186925, 3.6)
(0.190747, 3.7)
(0.194436,   3.8)
(0.197997, 3.9)
(0.201439, 4.)
(0.204767, 4.1)
(0.207987,   4.2)
(0.211106, 4.3)
(0.214128, 4.4)
(0.217058,  4.5)
(0.219901, 4.6)
(0.22266, 4.7)
(0.22534, 4.8)
(0.227945,  4.9)
(0.230478, 5.)
(0.232942, 5.1)
(0.235339, 5.2)
(0.237674,  5.3)
(0.239949, 5.4)
(0.242166, 5.5)
(0.244327, 5.6)
(0.246436,   5.7)
(0.248493, 5.8)
(0.250501, 5.9)
(0.252462, 6.)
(0.254378,   6.1)
(0.256251, 6.2)
(0.258081, 6.3)
(0.259871,  6.4)
(0.261622, 6.5)
(0.263335, 6.6)
(0.265012, 6.7)
(0.266654,   6.8)
(0.268263, 6.9)
(0.269838, 7.)
(0.271382, 7.1)
(0.272896,   7.2)
(0.27438, 7.3)
(0.275835, 7.4)
(0.277262, 7.5)
(0.278663,   7.6)
(0.280038, 7.7)
(0.281387, 7.8)
(0.282712,  7.9)
(0.284014, 8.)
};

\end{axis}

\end{tikzpicture}
  \caption{Lines of constant $c$ for different code rates - exhibiting similar behavior. From the left: the limit for the finite expected number of steps for unidirectional correction $c = -1$; for bidirectional correction $c = -\frac{1}{2}$; $c = -\frac{1}{4}$; $c = -\frac{1}{8}$; Shannon bound $c = 0$. The dashed lines represent the typical code rates.}
   \label{cval}
\end{figure*}
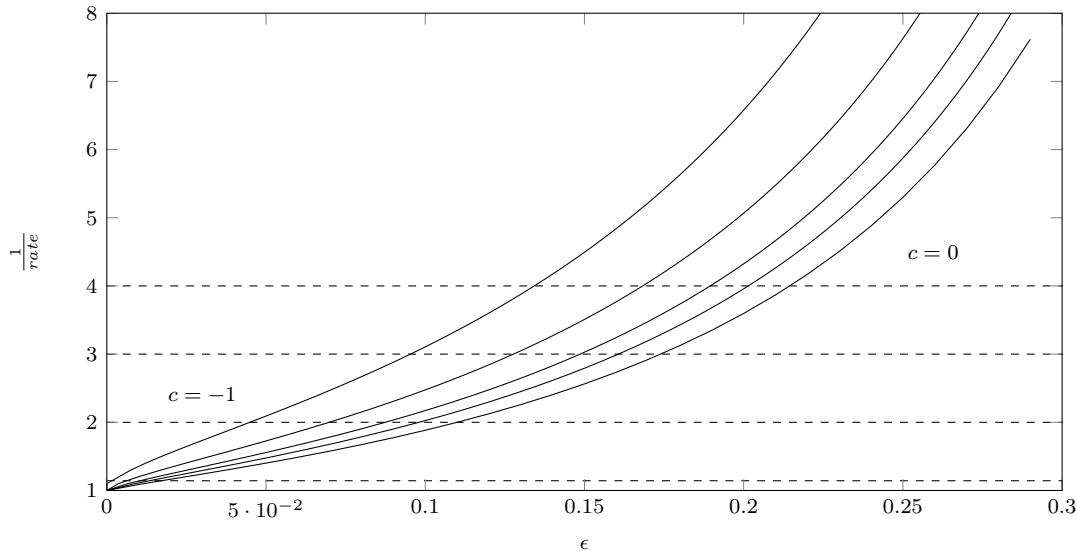

Having obtained the rate of the exponents $U(w)$ and $V(w)$ from (\ref{equv}) and (\ref{urel}), we can find the asymptotic behavior of the probability that a certain number of steps would be exceeded for a single node. Assume $U(w)\approx  c_u 2^{uw},\ V(w)\approx 1-c_v 2^{vw}$ for some unknown coefficients $c_u,\ c_v$. Now the asymptotic probability that the number of nodes of a wrong correction sub-tree for a given proper node will exceed a certain number of steps $s$ is:

$$\Pr\left(w>\frac{\lg(s/c_u)}{u}\right)\approx 1-c_v 2^{\frac{v}{u}\lg(s/c_u)}=
1-c_v \left(\frac{s}{c_u}\right)^{v/u}$$
\be\Pr(\#\textrm{ nodes in wrong subtree}>s)\approx 1-c_p s^c\ee

\noindent where $c:=v/u=1-1/u$ and $c_p=c_v/c_u^c$. Hence, we asymptotically obtain the Pareto probability distribution. The exponent, i.e. the shape of the distribution, can be found analytically. For code rate 1/2, the approximate values are:

\begin{table}[h]
\centering
\begin{tabular}{|c|c|c|c|c|c|c|c|c|}
  \hline   $\pb$ & 0.03 & 0.045 & 0.06 & 0.07 & 0.08 & 0.09 & 0.10 & 0.11 \\ \hline
  $c$ & -1.46 & -1 & -0.67 & -0.5 & -0.35 & -0.22 & -0.1 & 0\\
  \hline
\end{tabular}
\end{table}

This coefficient approaches zero in the Shannon limit ($\pb = 0.11$). For larger channel noises the weight function decreases statistically, therefore the weight drops would go to infinity. For lower channel noises, or if the errors were spread almost uniformly, the weight drops would remain close to zero, resulting in a small number of decoding steps.

The problem is that there can appear statistically large error concentrations which require a very large number of decoding steps for successful correction. If $c < -1$, their probability drops quicker than their influence, and the expected number of steps per node becomes finite. For $c\in [-1,0)$ it becomes infinite, but since we are working on finite data frames, we are still able to perform the correction with a high degree of probability.

The numbers of decoding steps per node cannot be treated independently. The surrounding nodes have similar weight drops, and large values would implicate a large number of steps in the whole area around the considered node. The number of decoding steps grows exponentially with the weight drop, so in practice most of the steps should be located around the nodes with the largest weight drops (error concentrations).

The amount of memory available to the decoder limits the number of steps that can be used to deal with an error concentration. A weight drop with a corresponding error concentration that cannot be corrected within a certain limit of steps is referred to as a CEC. The probability of not achieving this limit, i.e., that $s$ steps are insufficient to correct a $l$ node frame, is approximately

$$(1-c_p s^c)^l \approx 1-lc_p s^c$$

Increasing the number of steps a certain number of times, reduces the failure probability to approx. $c$-th power of this number of times. This allows us to reduce the failure probability to zero by simply increasing the number of the decoding steps. However, this is no longer practical near the Shannon limit. For example for $c=-0.1$, increasing the number of steps $2^{10}$ times would enable us to reduce the probability of failure only twofold. \\

A very high number of decoding steps increases the probability of system state collisions to a level where it cannot be safely ignored. The probability that two considered corrections up to a given symbol will correspond to the same system state is approximately proportional to the number of pairs, i.e. to the square of the width of the tree. However, the existence of such a collision does not necessarily mean that one of the two involved corrections will be used. The only real problem which can leave a few dozens of bits damaged is the collisions with the proper correction patterns. Their probability is proportional to the width of the tree. Such collisions can occur on each position, therefore this probability is approximately proportional to the total number of steps. For a 64 bit state and $5\cdot10^7$ steps it is practically negligible ($2.6\cdot 10^{-12}$) and it is difficult to imagine that, for example, a 128-bit state could turn out to be insufficient for any conceivable practical step limit. Below the Shannon limit, the size of the system state would need to grow logarithmically with the size of the frame, while for larger noises it would require to grow linearly to compensate for the lacking code rate.

\begin{figure*}[t!]
    \centering
        \input{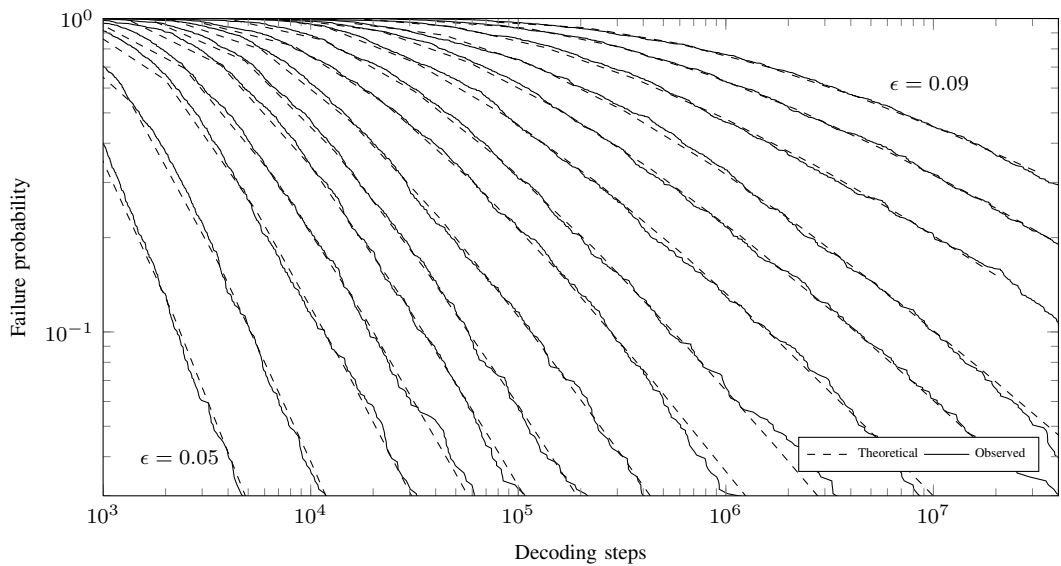}
        \caption{Experimental results of the approximate distribution function of the number of decoding steps required for 1024 byte frames, the $1/2$ rate (0.49) and channel BER successively 0.05, 0.055, 0.06, 0.0625, 0.065, 0.0675, 0.07, 0.0725, 0.075, 0.0775, 0.08, 0.0825, 0.085, 0.0875 and 0.09. In a log-log scale, it should approach the $2c$ direction - the dashed line is the result of fitting $A, B, C$ coefficients of $2cx+A+Be^{-Cx}$ equation.}
        \label{fig:theory-fit}
\end{figure*}

\subsection{Bidirectional Correction}

The unidirectional correction will stop on a single CEC. In this case, on average only half of the frame is fully corrected. With bidirectional correction, the CEC would be approached from both sides, which in most cases allows for successful correction. During our experimental evaluation, more than 80\% of such CECs were successfully corrected. Even if a single CEC cannot be dealt with, the data stream was successfully corrected up to the CEC from both sides, which means that only a few dozen known bits around the CEC remain unrepaired efficiently. As a result, for bidirectional correction, two CECs are required to essentially cripple the correction process. The probability that there will be at most one CEC is approx.:

$$(1-c_p s^c)^l+ lc_p s^c (1-c_p s^c)^{l-1}\approx 1- l^2 (c_p)^2 s^{2c}$$

The failure probability drops to approximately the square of the original, while the Pareto shape parameter approximately doubles. As a result, a ten-fold increase of the step limit reduces the probability of failure asymptotically $10^{-2c}$ number of times. For the 1/2 rate, the approximate values are:

\begin{table}[h]
\centering
\begin{tabular}{|c|c|c|c|c|c|c|c|c|}
  \hline   $\pb$ & 0.03 & 0.045 & 0.06 & 0.07 & 0.08 & 0.09 & 0.10 & 0.11 \\ \hline
  $10^{-2c}$ & 846 & 99 & 22 & 10 & 5 & 2.7 & 1.6 & 1 \\
  \hline
\end{tabular}
\end{table}

The probability of failure also drops reversely proportional to approximately the square of the length of the frame ($l$), therefore the best performance is expected for short data frames. However, since each frame requires the final system state to be transmitted, shortening the frame would quickly result in a severe reduction of the code rate. As a result, a compromise is required.

A convenient way of comparing the theoretical results with the simulations is to order them by the number of the required decoding steps. The position in this order divided by their total number approximates the probability of performing a successful correction within the given step limit. As a result, we obtain an approximate distribution function of the required number of steps. This also allows for extrapolation of the bit error rate for lower channel noises. Based on the theoretical analysis presented, in the log-log scale it should approach the direction determined by $2c$. From Fig.~\ref{fig:theory-fit} we see that it exponentially approaches it at an unknown rate. However, the asymptotic direction corresponds well with the theory.

For low channel error rates, we were not able to achieve valid BER estimates from the experiments, as all simulated frames were transmitted correctly in a feasible computation time. In these cases, we use the described fits. In order to estimate the number of uncorrected bits with the insufficient number of decoding steps, we use the above probabilities of failure to calculate the probability of a single CEC and thus determine the probability distribution for the number of CECs. If there are $m$ separate CECs, we should repair up to the first one from each side, what is on average approximately

$$\int_0^1 \int_0^1..\int_0^1 \min(x_1,..,x_m)\ dx_1..dx_m=\frac{1}{m+1}$$

\noindent of the whole frame; in this case, approx. $\frac{m-1}{m+1}$ bits should remain unrepaired.

\section{Experimental Evaluation}
\label{sec:evaluation}

This section describes the results of experimental evaluation of the proposed error correction method. We have implemented the described correction tree algorithm in C++. The sources are available at \cite{correction-trees-sources} under the GPL license.

The implemented codec is currently optimized for $\nn=8,\ \RR\in\{1,2,3,4,5,6\}$ and $\nn=9,\ \RR=6$ cases. However, it can be easily expanded to any case $\nn\in[2,16]$, $\RR\in [1,n-2]$ by adding corresponding optimized definitions of immediately produced redundancy bits. In this study, $\RR$ remains constant for all the blocks. Dynamic changes of $\RR$ are a straightforward modification of the proposed scheme and allow for designing codes with an arbitrary rate. These $\nn$-bit blocks can be grouped into frames of an arbitrary size, as shown in Section~\ref{sec:final-state}. In the experiments described, we use frame sizes $l$ of 1024 and 4096. We use $\nn = 8$ and $\RR = \lbrace 1, 4, 6 \rbrace$ to obtain code rates of 0.875, 0.5 and 0.25 respectively. For the rate of $1/3$ we use $\nn = 9$ and $\RR = 6$.

This study deals with reliable transmission over the BSC. The $\pb$ for the performed experiments has been carefully chosen for each case to cover the range where the codec ceases to provide reliable transmission capability. In each experiment, the codec transmits 1000 frames and collects the necessary statistics. Each frame is composed of a pseudo-random payload and the corresponding redundancy information appended by the encoder. The seed for the utilized pseudo-random number generator is changed on a per-frame basis.

The results obtained for the proposed correction method are presented and discussed in Section~\ref{sec:setup}. A comparison with existing error correction codes is presented in Section~\ref{sec:comparison}.

\subsection{Correction Tree Efficiency}
\label{sec:setup}

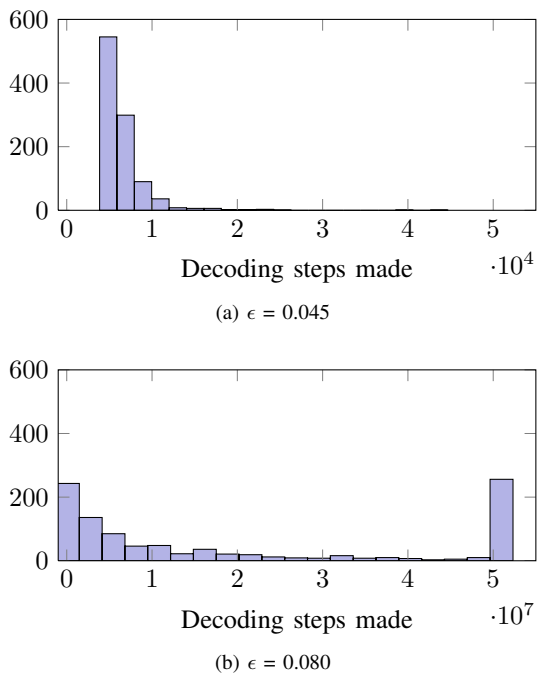
\begin{figure}[t]
 \centering
  \subfloat[$\pb$ = 0.045]{\begin{tikzpicture}

% defining custom colors
\definecolor{mycolor1}{rgb}{0.7,0.7,0.9}

% Axis at [0.13 0.19 0.78 0.69]
\begin{axis}[
scale only axis,
width=2.5in,
height=1in,
xmin=-1000, xmax=55000,
ymin=0, ymax=600,
xlabel={Decoding steps made},
axis on top]

\addplot[ybar,bar width=0.09in, bar shift=0in,fill=mycolor1,draw=black] plot coordinates{ 
(4862.0, 545)
 (6904.25, 299)
 (8946.5, 90)
 (10988.75, 36)
 (13031.0, 8)
 (15073.25, 6)
 (17115.5, 6)
 (19157.75, 2)
 (21200.0, 2)
 (23242.25, 3)
 (25284.5, 1)
 (27326.75, 0)
 (29369.0, 0)
 (31411.25, 0)
 (33453.5, 0)
 (35495.75, 0)
 (37538.0, 0)
 (39580.25, 1)
 (41622.5, 0)
 (43664.75, 1)
};

\end{axis}

\end{tikzpicture}} \hspace{0.5cm}
  \subfloat[$\pb$ = 0.080]{\begin{tikzpicture}

% defining custom colors
\definecolor{mycolor1}{rgb}{0.7,0.7,0.9}

% Axis at [0.13 0.19 0.78 0.69]
\begin{axis}[
scale only axis,
width=2.5in,
height=1in,
xmin=-1000000, xmax=55000000,
ymin=0, ymax=600,
xlabel={Decoding steps made},
axis on top]

\addplot[ybar,bar width=0.12in, bar shift=0in,fill=mycolor1,draw=black] plot coordinates{ 
 (130127.0, 243)
 (2806231.9500000002, 136)
 (5482336.9000000004, 85)
 (8158441.8500000006, 46)
 (10834546.800000001, 48)
 (13510651.75, 22)
 (16186756.700000001, 36)
 (18862861.650000002, 21)
 (21538966.600000001, 19)
 (24215071.550000001, 12)
 (26891176.5, 9)
 (29567281.450000003, 8)
 (32243386.400000002, 16)
 (34919491.350000001, 8)
 (37595596.300000004, 10)
 (40271701.25, 7)
 (42947806.200000003, 3)
 (45623911.150000006, 5)
 (48300016.100000001, 10)
 (50976121.050000004, 256)
};

\end{axis}

\end{tikzpicture}}
  \caption{The number of decoding steps performed by the decoder for the rate 1/2 code and $l = 4096$ blocks}
  \label{fig:steps-histograms}
\end{figure}

The most important criterion for the efficiency of the proposed correction method is the number of the necessary decoding steps. The limit of the decoding steps, which stems from the available computational resources directly affects the correction performance. If the transmission channel does not generate any errors, the decoder performs only $l$ steps. More decoding steps are needed when transmission errors occur. Figure~\ref{fig:steps-histograms} shows the histogram of the necessary decoding steps for $l = 4096$ and two example channels: $\pb = 0.045$ and $\pb = 0.08$. In case of the latter, the visible peak at the end of the histogram represents the frames damaged beyond repair. The remaining 75\% of the frames have been successfully corrected with 10,367,638 decoding steps on average. The results from all of the performed experiments are collected in Table~\ref{tab:results}.

% Preview source code for paragraph 0
\begin{table*}[t]
\caption{The achieved frame errors (FE) and the average number of decoding steps $\overline{D}$ per symbol for the performed experiments (1 for pure decoding). The total number of simulated frames per setting is 1000.}
\label{tab:results}
\centering
\begin{tabular}{lrrrrrrrrl}
\toprule
\textbf{Rate} $\frac{7}{8}$ & 0.002 & 0.003 & 0.004 & 0.005 & 0.006 & 0.007 & 0.008 & 0.009 & $\pb$\tabularnewline
\midrule
\multirow{2}{*}{$l=1024$} & 0 & 0 & 0 & 0 & 0 & 1 & 0 & 14 & $FE$\tabularnewline
 & 1.274 & 1.702 & 2.882 & 5.447 & 18.76 & 133.9 & 232.6 & 1349 & $\overline{D}/l$\tabularnewline
\midrule
\multirow{2}{*}{$l=4096$} & 0 & 0 & 0 & 0 & 1 & 12 & 38 & 150 & $FE$\tabularnewline
 & 1.307 & 1.926 & 4.279 & 17.08 & 61.65 & 401.9 & 1199 & 3205 & $\overline{D}/l$\tabularnewline
\toprule
\textbf{Rate} $\frac{1}{2}$ & 0.05 & 0.06 & 0.065 & 0.07 & 0.075 & 0.08 & 0.085 & 0.09 & $\pb$\tabularnewline
\midrule
\multirow{2}{*}{$l=1024$} & 0 & 0 & 0 & 0 & 5 & 25 & 92 & 273 & $FE$\tabularnewline
 & 2.385 & 8.946 & 21.80 & 76.92 & 680.7 & 2660 & 8232 & 18795 & $\overline{D}/l$\tabularnewline
\midrule
\multirow{2}{*}{$l=4096$} & 0 & 0 & 1 & 14 & 57 & 229 & 659 & 945 & $FE$\tabularnewline
 & 3.134 & 17.36 & 111.5 & 510.1 & 1776 & 4678 & 9645 & 11949 & $\overline{D}/l$\tabularnewline
\toprule
\textbf{Rate} $\frac{1}{4}$ & 0.13 & 0.14 & 0.15 & 0.16 & 0.165 & 0.17 & 0.175 & 0.18 & $\pb$\tabularnewline
\midrule
\multirow{2}{*}{$l=1024$} & 0 & 0 & 0 & 0 & 0 & 0 & 5 & 16 & $FE$\tabularnewline
 & 1.873 & 2.988 & 7.972 & 21.12 & 89.90 & 193.7 & 705.0 & 2070 & $\overline{D}/l$\tabularnewline
\midrule
\multirow{2}{*}{$l=4096$} & 0 & 0 & 0 & 1 & 4 & 9 & 45 & 220 & $FE$\tabularnewline
 & 2.204 & 4.119 & 12.25 & 71.91 & 221.6 & 636.1 & 1658 & 4480 & $\overline{D}/l$\tabularnewline
\bottomrule
\end{tabular}
\end{table*}

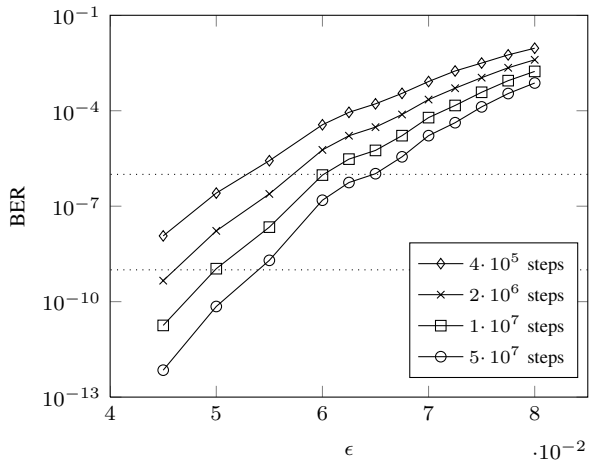
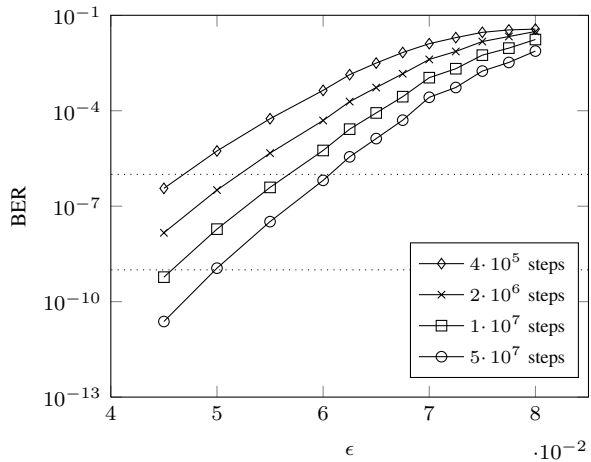
\begin{figure*}[t!]
 \centering
  \subfloat[$l = 1024$]{% This file was created by matlab2tikz v0.0.5.
% Copyright (c) 2008--2010, Nico Schlömer <nico.schloemer@ua.ac.be>
% All rights reserved.
%
% The latest updates can be retrieved from
%  http://win.ua.ac.be/~nschloe/content/matlab2tikz/
% and
%  http://www.mathworks.com/matlabcentral/fileexchange/22022 .
% where you can also make suggestions and rate matlab2tikz.

\begin{tikzpicture}[font=\footnotesize]

% Axis at [0.13 0.11 0.78 0.79]
\begin{semilogyaxis}[
scale only axis,
width=2.5in,
height=2in,
xmin=0.04, xmax=0.085,
ymin=1e-13, ymax=0.1,
xlabel={$\pb$},
ylabel={BER},
axis on top,
legend entries={$4 \cdotp 10^{5}$ steps, $2 \cdotp 10^{6}$ steps, $1 \cdotp 10^{7}$ steps, $5 \cdotp 10^{7}$ steps},
legend style={legend columns=1,at={(0.98,0.05)},anchor=south east,font=\scriptsize}]

\addplot [
color=black,
solid,
mark=diamond,
mark options={solid}
]
coordinates{
 (0.045, 1.16574e-08)
(0.05, 2.58325e-07)
(0.055, 2.69079e-06)
(0.06, 3.62477e-05)
(0.0625, 8.91015e-05)
(0.065, 0.000164609)
(0.0675, 0.000351882)
(0.07, 0.000827926)
(0.0725, 0.00177607)
(0.075, 0.00315378)
(0.0775, 0.00569819)
(0.08, 0.00922949)
};

\addplot [
color=black,
solid,
mark=x,
mark options={solid}
]
coordinates{
(0.045, 4.56967e-10)
(0.05, 1.67535e-08)
(0.055, 2.43183e-07)
(0.06, 5.87104e-06)
(0.0625, 1.63933e-05)
(0.065, 3.05293e-05)
(0.0675, 7.61146e-05)
(0.07, 0.000224357)
(0.0725, 0.000510005)
(0.075, 0.00109555)
(0.0775, 0.00224164)
(0.08, 0.0039877)
};

\addplot [
color=black,
solid,
mark=square,
mark options={solid}
]
coordinates{
(0.045, 1.79131e-11)
(0.05, 1.08654e-09)
(0.055, 2.1978e-08)
(0.06, 9.50933e-07)
(0.0625, 3.01611e-06)
(0.065, 5.66212e-06)
(0.0675, 1.64641e-05)
(0.07, 6.07978e-05)
(0.0725, 0.00014645)
(0.075, 0.000380569)
(0.0775, 0.000881853)
(0.08, 0.00172292)
};

\addplot [
color=black,
solid,
mark=o,
mark options={solid}
]
coordinates{
(0.045, 7.0219e-13)
(0.05, 7.04667e-11)
(0.055, 1.98629e-09)
(0.06, 1.54023e-07)
(0.0625, 5.54919e-07)
(0.065, 1.05013e-06)
(0.0675, 3.56131e-06)
(0.07, 1.64754e-05)
(0.0725, 4.20535e-05)
(0.075, 0.000132201)
(0.0775, 0.000346917)
(0.08, 0.000744407)
};

\addplot [
color=black,
dotted
]
coordinates{
 (0.0,1e-6) (0.1,1e-6)
};

\addplot [
color=black,
dotted
]
coordinates{
 (0.0,1e-9) (0.1,1e-9)
};

\end{semilogyaxis}

\end{tikzpicture}} \hspace{0.5cm}
  \subfloat[$l = 4096$]{% This file was created by matlab2tikz v0.0.5.
% Copyright (c) 2008--2010, Nico Schlömer <nico.schloemer@ua.ac.be>
% All rights reserved.
%
% The latest updates can be retrieved from
%  http://win.ua.ac.be/~nschloe/content/matlab2tikz/
% and
%  http://www.mathworks.com/matlabcentral/fileexchange/22022 .
% where you can also make suggestions and rate matlab2tikz.

\begin{tikzpicture}[font=\footnotesize]

% Axis at [0.13 0.11 0.78 0.79]
\begin{semilogyaxis}[
scale only axis,
width=2.5in,
height=2in,
xmin=0.04, xmax=0.085,
ymin=1e-13, ymax=0.1,
xlabel={$\pb$},
ylabel={BER},
axis on top,
legend entries={$4 \cdotp 10^{5}$ steps, $2 \cdotp 10^{6}$ steps, $1 \cdotp 10^{7}$ steps, $5 \cdotp 10^{7}$ steps},
legend style={legend columns=1,at={(0.98,0.05)},anchor=south east,font=\scriptsize}]

\addplot [
color=black,
solid,
mark=diamond,
mark options={solid}
]
coordinates{
(0.045, 3.63755e-7)
(0.05, 5.444e-6)
(0.055,    0.000056081)
(0.06, 0.000437183)
(0.0625, 0.0013602)
(0.065,    0.0031467)
(0.0675, 0.00677784)
(0.07, 0.0128102)
(0.0725,    0.0198302)
(0.075, 0.0291227)
(0.0775, 0.0346546)
(0.08,    0.0367715)
};

\addplot [
color=black,
solid,
mark=x,
mark options={solid}
]
coordinates{
(0.045, 1.46289e-8)
(0.05, 3.2093e-7)
(0.055,    4.65433e-6)
(0.06, 0.0000493876)
(0.0625,    0.000193273)
(0.065, 0.000535189)
(0.0675, 0.00145258)
(0.07,    0.0041018)
(0.0725, 0.00729447)
(0.075, 0.0149651)
(0.0775,    0.0215056)
(0.08, 0.0309786)
};

\addplot [
color=black,
solid,
mark=square,
mark options={solid}
]
coordinates{
(0.045, 5.89243e-10)
(0.05,    1.90112e-8)
(0.055, 3.89575e-7)
(0.06,    5.63892e-6)
(0.0625, 0.0000263793)
(0.065,    0.0000852876)
(0.0675, 0.000276637)
(0.07, 0.00109422)
(0.0725,    0.00209222)
(0.075, 0.00558285)
(0.0775, 0.0092903)
(0.08,    0.0173289)
};

\addplot [
color=black,
solid,
mark=o,
mark options={solid}
]
coordinates{
(0.045, 2.37419e-11)
(0.05,    1.12758e-9)
(0.055, 3.27115e-8)
(0.06,    6.48154e-7)
(0.0625, 3.55651e-6)
(0.065,    0.000013321)
(0.0675, 0.0000505285)
(0.07, 0.00026454)
(0.0725,    0.000538175)
(0.075, 0.00176395)
(0.0775, 0.00329675)
(0.08,    0.00750733)
};

\addplot [
color=black,
dotted
]
coordinates{
 (0.0,1e-6) (0.1,1e-6)
};

\addplot [
color=black,
dotted
]
coordinates{
 (0.0,1e-9) (0.1,1e-9)
};

\end{semilogyaxis}

\end{tikzpicture}}
  \caption{Performance of the correction tree algorithm for a $1/2$ rate code and different step limits. The dotted lines represent the commonly used thresholds for reliable transmission, i.e. BER of $10^{-6}$ and $10^{-9}$.}
  \label{fig:performance-step-limits}
\end{figure*}

The proposed algorithm allows for correction arbitrarily close to the Shannon limit. The only limitation is the amount of resources that are available on the receiving end of the transmission system. The maximum number of possible decoding steps stems directly from the available system memory and computational power. The size of the decoding node structure is 40B, and thus the memory requirements for step limits of $4 \cdotp 10^{5}$, $2 \cdotp 10^{6}$, $1 \cdotp 10^{7}$ and $5 \cdotp 10^{7}$ are 15MB, 76MB, 381MB and 1.9GB, respectively. At the cost of the decoding performance, it is possible to reduce the size of the correction tree node to approximately 20B. Figure~\ref{fig:performance-step-limits} shows the achievable transmission error rates for these step limits for $l = 1024$ and $l = 4096$. On a standard PC with a single 3.16~GHz core, the correction process is performed at a speed of approximately 3.2 million steps per second. For smaller channel noise the processing speed reaches up to approx. 5 million steps per second. For larger ones it drops to an average 2 million steps per second. Combining these speeds with average steps per symbol from Table~\ref{tab:results}, we see that this software implementation of the proposed approach is capable of processing up to a few megabytes per second on average for low noise levels. The processing speed drops gradually to a few hundred bytes per second, for extreme noise levels.

\subsection{Comparative Evaluation}
\label{sec:comparison}

We compared the proposed correction tree algorithm with existing state-of-the-art forward error correction codes, namely the LDPC \cite{gallager,mackay} and turbo codes \cite{berrou}. We implemented a BSC simulation scenario in C++ using the IT++ library \cite{itpp}. The soft-information for the decoder input was calculated taking into account the log likelihood ratio (LLR) for the BSC channel, i.e. $log_2\frac{\pb}{1-\pb}$.

In this experiment, we consider two most popular coding rates: $\frac{1}{2}$ and $\frac{1}{3}$. The utilized LDPC codes were either generated randomly or taken from \cite{ldpc-encyclopedia}. We used the belief propagation decoder. A summary of the LDPC codes used is shown in Table~\ref{tab:ldpc}.

For the turbo code, we used the (13,15) code from the Wideband Code Division Multiple Access (WCDMA) standard and the maximum a-posteriori probability (MAP) decoder with 10 iterations \cite{itpp}.

\begin{figure*}[t]
 \centering
  \subfloat[Code rate $\frac{1}{2}$ (Shannon limit is 0.11)]{% This file was created by matlab2tikz v0.0.5.
% Copyright (c) 2008--2010, Nico Schlömer <nico.schloemer@ua.ac.be>
% All rights reserved.
%
% The latest updates can be retrieved from
%  http://win.ua.ac.be/~nschloe/content/matlab2tikz/
% and
%  http://www.mathworks.com/matlabcentral/fileexchange/22022 .
% where you can also make suggestions and rate matlab2tikz.

\begin{tikzpicture}[font=\footnotesize]

% Axis at [0.13 0.11 0.78 0.79]
\begin{semilogyaxis}[
scale only axis,
width=2.5in,
height=2in,
xmin=0.035, xmax=0.0925,
ymin=1e-7, ymax=0.5,
xlabel={$\pb$},
ylabel={BER},
axis on top,
legend entries={LDPC MK 4000, LDPC MK 20000, LDPC R 500, CT 1024B},
legend style={legend columns=2,at={(0.98,0.05)},anchor=south east,font=\tiny}]

\addplot [
color=black,
dashed,
mark=diamond,
mark options={solid}
]
coordinates{
(0.1, 0.094125)
(0.095, 0.086875)
(0.09, 0.0626389)
(0.085, 0.0571389)
(0.08, 0.02053)
(0.075, 0.00892623)
(0.07, 0.000618103)
(0.065, 3.34617e-05)
(0.06, 9.13043e-07)
};

\addplot [
color=black,
dashed,
mark=x,
mark options={solid}
]
coordinates{
(0.1, 0.10235)
(0.095, 0.081875)
(0.09, 0.073475)
(0.085, 0.063)
(0.08, 0.025325)
(0.075, 0.000628481)
(0.07, 2.052e-07)
};

\addplot [
color=black,
dashed,
mark=square,
mark options={solid}
]
coordinates{
(0.1, 0.0892609)
(0.095, 0.0656066)
(0.09, 0.0647419)
(0.085, 0.0509)
(0.08, 0.032189)
(0.075, 0.0202412)
(0.07, 0.0134047)
(0.065, 0.00687838)
(0.06, 0.00312812)
(0.055, 0.00103208)
(0.05, 0.000379203)
(0.045, 0.000110062)
(0.04, 1.83814e-05)
};

\addplot [
color=black,
solid,
mark=triangle,
mark options={solid}
]
coordinates{
(0.045, 7.0219e-13)
(0.05, 7.04667e-11)
(0.055, 1.98629e-09)
(0.06, 1.54023e-07)
(0.0625, 5.54919e-07)
(0.065, 1.05013e-06)
(0.0675, 3.56131e-06)
(0.07, 1.64754e-05)
(0.0725, 4.20535e-05)
(0.075, 0.000132201)
(0.0775, 0.000346917)
(0.08, 0.000857)
(0.0825,0.00263)
(0.085, 0.00568)
(0.0875, 0.00818)
(0.09,0.0166)
(0.0925,0.0289)
(0.095,0.0509)
%(0.08, 0.000744407)
%(0.0825,0.0010938)
%(0.085, 0.0027257)
%(0.0875, 0.00519654)
%(0.09,0.0100549)
%(0.0925,0.0180562)
%(0.095,0.0291383)
};

\addplot [
color=black,
dotted
]
coordinates{
 (0.0,1e-6) (0.2,1e-6)
};

\addplot [
color=black,
dotted
]
coordinates{
 (0.0,1e-9) (0.2,1e-9)
};

\end{semilogyaxis}

\end{tikzpicture}} \hspace{0.5cm}
  \subfloat[Code rate $\frac{1}{3}$ (Shannon limit is 0.174)]{% This file was created by matlab2tikz v0.0.5.
% Copyright (c) 2008--2010, Nico Schlömer <nico.schloemer@ua.ac.be>
% All rights reserved.
%
% The latest updates can be retrieved from
%  http://win.ua.ac.be/~nschloe/content/matlab2tikz/
% and
%  http://www.mathworks.com/matlabcentral/fileexchange/22022 .
% where you can also make suggestions and rate matlab2tikz.

\begin{tikzpicture}[font=\footnotesize]

% Axis at [0.13 0.11 0.78 0.79]
\begin{semilogyaxis}[
scale only axis,
width=2.5in,
height=2in,
xmin=0.09, xmax=0.16,
ymin=1e-7, ymax=0.5,
xlabel={$\pb$},
ylabel={BER},
axis on top,
legend entries={WCDMA TC 960b, WCDMA TC 9216b, LDPC MK 1920b, CT 9216b},
legend style={legend columns=2,at={(0.98,0.05)},anchor=south east,font=\tiny}]

\addplot [
color=black,
densely dashed,
mark=diamond,
mark options={solid}
]
coordinates{
(0.25, 0.440833)
(0.245, 0.489183)
(0.24, 0.464509)
(0.235, 0.474107)
(0.23, 0.454911)
(0.225, 0.434375)
(0.22, 0.44375)
(0.215, 0.432292)
(0.21, 0.447098)
(0.205, 0.432708)
(0.2, 0.39082)
(0.195, 0.434375)
(0.19, 0.378493)
(0.185, 0.37261)
(0.18, 0.353993)
(0.175, 0.348355)
(0.17, 0.322656)
(0.165, 0.301935)
(0.16, 0.246875)
(0.155, 0.195644)
(0.15, 0.154192)
(0.145, 0.150818)
(0.14, 0.079193)
(0.135, 0.0534664)
(0.13, 0.0203275)
(0.125, 0.00954785)
(0.12, 0.00493002)
(0.115, 0.00107629)
(0.11, 0.000307337)
(0.105, 8.98931e-05)
(0.1, 2.79397e-05)
};

\addplot [
color=black,
densely dashed,
mark=x,
mark options={solid}
]
coordinates{
(0.17, 0.346354)
(0.165, 0.308919)
(0.16, 0.300515)
(0.155, 0.254858)
(0.15, 0.230165)
(0.145, 0.116432)
(0.14, 0.0260964)
(0.135, 0.00282755)
(0.13, 6.77958e-05)
};

\addplot [
color=black,
dashed,
mark=square,
mark options={solid}
]
coordinates{
(0.2, 0.202564)
(0.195, 0.187537)
(0.19, 0.183854)
(0.185, 0.179975)
(0.18, 0.169708)
(0.175, 0.152515)
(0.17, 0.149103)
(0.165, 0.143961)
(0.16, 0.134141)
(0.155, 0.108142)
(0.15, 0.104542)
(0.145, 0.0766008)
(0.14, 0.0569276)
(0.135, 0.0281446)
(0.13, 0.0172573)
(0.125, 0.0086018)
(0.12, 0.00248785)
(0.115, 0.000627742)
(0.11, 0.000125594)
(0.105, 1.46713e-05)
(0.1, 1.9856e-06)
(0.095, 2.284e-07)
};

\addplot [
color=black,
solid,
mark=triangle,
mark options={solid}
]
coordinates{
(0.1, 5.27937e-11)
(0.11, 9.84437e-9)
(0.12,7.78172e-7)
(0.13, 0.0000385197)
(0.135, 0.000228731)
(0.1375, 0.000552927)
(0.14, 0.00115287)
(0.1425, 0.00212257)
(0.145, 0.0048823)
(0.1475, 0.00882076)
(0.15, 0.0186)
(0.1525, 0.0319)
(0.155, 0.0502)
(0.1575, 0.0690)
(0.16, 0.0943)
%(0.145, 0.0048823)
%(0.1475, 0.00882076)
%(0.15, 0.0143341)
%(0.1525, 0.0243426)
%(0.155, 0.036632)
%(0.1575, 0.0461631)
%(0.16, 0.056899)
};

\addplot [
color=black,
dotted
]
coordinates{
 (0.0,1e-6) (0.2,1e-6)
};

\addplot [
color=black,
dotted
]
coordinates{
 (0.0,1e-9) (0.2,1e-9)
};

\end{semilogyaxis}

\end{tikzpicture}}
  \caption{Correction performance comparison with popular LDPC and turbo codes on a binary symmetric channel.}
  \label{fig:comparison}
\end{figure*}
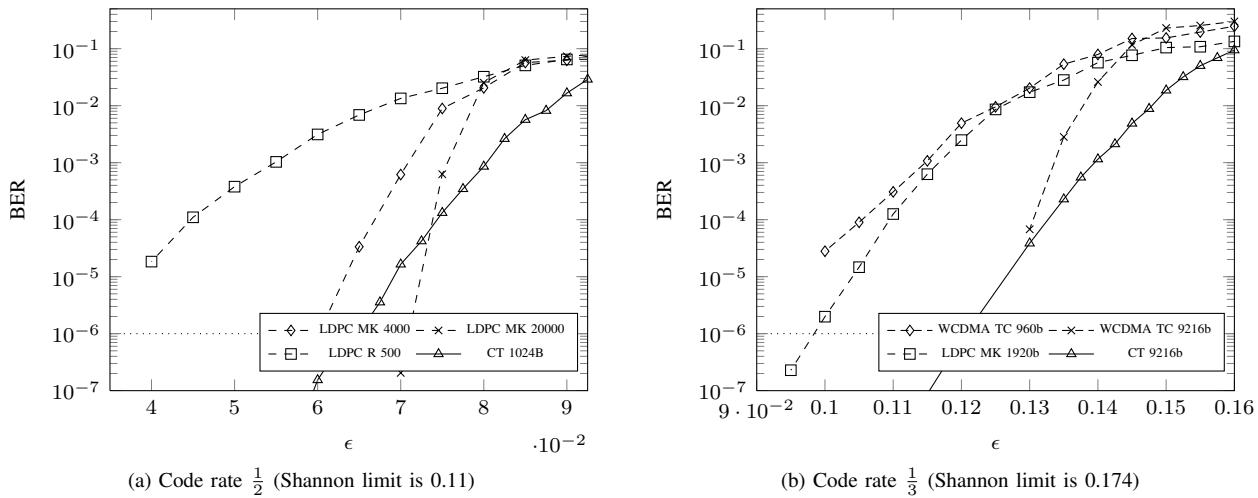

The results are shown in Fig.~\ref{fig:comparison}. The proposed correction algorithm delivers a very good correction performance. It outperforms the commonly available LDPC and turbo codes for small frame sizes and for higher channel error rates.

\begin{table}[t]
 \caption{The utilized LDPC codes}
 \label{tab:ldpc}
 \centering
 \begin{tabular}{lrrrr}
  \toprule
  \textbf{Code} & \textbf{l} & \textbf{K} & \textbf{Rate} & \textbf{Origin} \\
  \midrule
  MK 1920 & 1920 & 1280 & $1/3$ & 1920.1280.3.303 \cite{ldpc-encyclopedia} \\
  MK 4000 & 4000 & 2000 & $1/2$ & 4000.2000.3.243~\cite{ldpc-encyclopedia} \\
  MK 20000 & 20000 & 10000 & $1/2$ & 20000.10000.3.631 \cite{ldpc-encyclopedia} \\
  R 500 & 500 & 250 & $1/2$ & Randomly generated using \cite{itpp} \\
  \bottomrule
 \end{tabular}
\end{table}

\section{Further Perspectives}
\label{sec:perspectives}

The discussed correction process is usually very fast for practical settings and does not require large amounts of memory; however, with probability decreasing to zero, the amount of required resources increases to infinity (Fig. \ref{fig:theory-fit}). It suggests that high capacity decoders should be designed in a hierarchical way, where the stream is processed by many low memory decoders first. The frames would be passed to more powerful decoders only if they have failed.

In this section, we discuss the construction of extremely large memory decoders which are not limited by the amount of random access memory (RAM) on a single computer. We also discuss a generalization of the correction process for different communication channels. Specifically, we consider the erasure channel, the additive white Gaussian noise (AWGN) channel, and channels with synchronization errors.

\subsection{Handling Extremely Large Correction Trees}

The presented algorithm was optimized for single core processing and within the capacity of RAM. In extreme situations, it may be necessary to allow for more decoding steps to increase the correction performance. For example using, a thousand times more steps for the 1/2 rate and $\pb=0.09$ would reduce the probability of failure about 20 fold. This section discusses the potential of generalizing the correction procedure to allow for efficient use of memory swapping or processing on multiple computers.

The original algorithm uses 3 large data structures: for storing the correction tree, the heap, and the binary search trees for each symbol position. Remembering the tree structure is useful for rapid application of the obtained correction path at a later stage, although it requires a lot of memory. While we would like to exchange data efficiently with the swapped memory segments or other computers, referring to such tables of nodes becomes problematic. However, the correction path can be retraced even without this structure. Having the proper state for any given position and a list of used states for the previous position, we can try successive corrections to make a step in the reverse direction until we get to a state from this list and so on.

If we no longer need the tree structure, the nodes can be handled independently, e.g. on different computers. The required information to be taken from the heap includes the symbol position, the system state before applying the correction, the currently considered correction of a given symbol, and the weight after applying this correction (in total, approx. 16B per tree node). In a single step, such a node would be retrieved from the heap and returned there with the updated correction pattern and the corresponding weight. Finally, the part of such a heap below a certain weight limit can be swapped for prospective consideration at a processing stage later.

When using multiple computers, each can have its own heap. Every node goes to only one such heap. Without communication, it is expected that the difference between the maximum weights of these heaps would increase. In order to prevent it, every few steps the considered nodes should be sent to the heap which has the lowest maximum weight in the cluster, instead of being stored in the local heap.

Additionally, we need to optimize the storage of the lists of the considered states for each symbol position. Originally used for stitching the two correction directions only, it would now also be used for retracing the proper correction path. We replace the previously used binary search trees with B-trees. They can be seen as a generalization, in which the nodes of the structure not only contain the state information and two pointers to its children, but become larger data blocks corresponding to a certain range of system states. Such a node contains a sorted list of system states from this range and potentially a sorted list of pointers to nodes corresponding to the sub-ranges. When a node exceeds its capacity, it is split into two parts and the structure of the B-tree needs to be updated. Using these large nodes, it is possible to insert a state or check if it is in the list by accessing only a few such data blocks. Therefore, such operations remain relatively fast, even when they are spread across the swapped memory or over the processing cluster.

In summary, working with extremely large trees involves restricting the operation of independent correction processors to the heaps which can be developed independently. Practical implementation of such a scheme would require a lot of work, but it may be useful for offline correction of critical data.

\subsection{Other Channels and Applications}

In this study, we have focused on the BSC. This section briefly discusses a generalization of the presented correction approach to different types of communication channels.

In the erasure channel, the received bits are either certain or completely lost. As such, this channel can be handled by sequential decoding, with the choice of $\pb=0$ for undamaged bits and $\pb=1/2$ for the erased ones. The fact that all possible corrections up to a given symbol position are equally probable allows for a simpler correction approach, i.e. to store all corrections which pass the verification up to a given position, expand them to the succeeding position and so on. Denoting bit erasure probability by $p_e$, the analogous theoretical Pareto coefficient (\ref{erasc}) for uni-directional correction and for $1/2$ rate are (derivation in Appendix):

\begin{table}[h]
\centering
\begin{tabular}{|c|c|c|c|c|c|c|c|c|}
 \hline   $p_e$ & 0.4 & 0.42 & 0.44 & 0.46 & 0.47 & 0.48 & 0.49 & 0.5 \\ \hline
 $c_e$ & -1.17 & -0.93 & -0.7 & -0.46 & -0.34 & -0.23 & -0.11 & 0 \\
 \hline
\end{tabular}
\end{table}

Fountain codes are a different class of erasure codes. They can produce an arbitrarily large number of data blocks, such that their undamaged sufficient portion allows us to reconstruct the original message. Such $i^{th}$ data block can be constructed by concatenation of $i^{th}$ redundancy bits from all symbols. This approach allows us to see it as an erasure channel with damages distributed in completely uniform way - without error concentrations and therefore with much better performance. Using correction trees would additionally allow us to retrieve almost all of the information also from damaged data blocks, which need to be ignored in standard fountain codes.

The presented algorithm can also be easily adapted to communication channels with $\pb$ being variable on a per bit basis, e.g. the AWGN channel. It is straightforward to include soft decoding information while calculating the weights of the successive nodes (\ref{eq:weight}). Its technical inconvenience is that it can change the order of corrections for a single symbol, which slightly complicates the algorithm. Theoretical analysis in this case requires replacing equations such as (\ref{vvrel}) and (\ref{urel}) with corresponding integral equations.

The proposed sequential decoding can also handle different types of local errors by simply adding new types of child nodes. For example synchronization errors, such as bit deletion or duplication, which are difficult to handle with other correction methods. The previously considered BSC set of symbol correction patterns is a set of masks for appropriate bit flips. Consideration of bit deletion, for example, involves expanding the set of correction patterns to include all possible bit deletions with the weight decreased by a logarithm of deletion probability, correspondingly shifting succeeding symbols. It is also possible to detect and undo global bijective transformations from some assumed family. When simultaneously considering the same transformation with different parameters, the weights of the proper settings should quickly dominate the others. The initial weights should be chosen as minus logarithm of transformation probability.
\section{Conclusions}
\label{sec:conclusions}

In this study, we presented a new approach to error correction using correction trees for efficient selection of possible correction patterns. The main contributions of our work include:

\begin{itemize}
 \item using a much larger system state space and a new efficient coding paradigm optimized for this purpose,
 \item utilization of a bidirectional correction mechanism, which greatly reduces the probability of failure,
 \item adoption of a heap for correction pattern selection, which allows for logarithmic time in large tree access operations.
\end{itemize}

This study also describes numerous optimizations which allows for efficient practical implementation of the proposed correction approach. Our correction algorithm allows for practically complete correction with negligible processing time, as long as the channel error rate is lower than approx. half (for code rate 1/2) the Shannon theoretical limit. For larger error rates, complete correction is still possible, but its cost in terms of computational complexity and memory requirements starts to increase rapidly. Commonly available error correction codes, such as LDPC or TC, are relatively costly for low error rates, and become unreliable as the channel noise increases, especially for small block sizes.

Due to the varying latency, the proposed method may not be not well suited for real-time applications like audio or video calls. However, there are certain classes of applications which could benefit from the proposed correction approach:

\begin{itemize}
 \item data storage - usually working on low error rate levels for which such correction is practically cost-free; however, time and random events can degrade the data, which could still be corrected provided that the degradation level is below the Shannon limit,
 \item far-space or underwater communication, which often transmits low amounts of data and where high correction cost can be easily afforded,
 \item authentication and reconstruction of digital content,
 \item applications where the redundancy should be used by authorized recipients only,
 \item communication channels with possible synchronization errors, like bit deletion or duplication.
\end{itemize}

% \appendix
\section*{Appendix}
\begin{figure*}[!t]
\normalsize
\setcounter{MYtempeqncnt}{\value{equation}}
\setcounter{equation}{19}
$$ T(s):=\textrm{probability that the number of corrections per symbol is at most }2^s  $$
\be T(s)=\left\{\begin{array}{ll} p_e T(s-\lg(\tilde{p_d})-1)+(1-p_e)T(s-\lg(\tilde{p_d}))\quad & \textrm{for }s\geq 0\\
                 0 & \textrm{for }s<0 \end{array}\right. \label{ereq}\ee
\setcounter{equation}{\value{MYtempeqncnt}}
\hrulefill
\vspace*{4pt}
\end{figure*}

We will now find Pareto coefficients for the case of the erasure channel: we can be sure of bits in some positions, although the remaining ones are lost completely. Let us denote this probability of erasure by $p_e$. For a $J=j\nn$ bit long message, on average $Jp_e$ bits are damaged; this means that $2^{Jp_e}$ possibilities remain, from which $(1-p_d)^j$ of the incorrect ones will survive the redundancy bit checks:
$$ 2^{j\nn p_e}(1-p_d)^j=\left(2^{\nn p_e+\lg(\tilde{p_d})}\right)^j=$$
\be\left(2^{\nn p_e+\kk-\nn}\right)^j=\left(2^{\kk-\nn(1-p_e)}\right)^j \label{eras}\ee
therefore we correct trees faster than they grow for $\frac{\kk}{\nn}<1-p_e$, which is Shannon capacity again.

By choosing $\pb=1/2$ for the lost bits, we could use sequential correction as previously. However, since there is no distinction between the allowed corrections of a given length, this time we can use a simpler and less demanding algorithm: remember all allowed corrections up to the given point, use such a list to find one for succeeding position and so on. The memory limit restricts the maximum number of nodes for a single position; with a nonzero probability, any such limit could be insufficient. Next we will find an analogous coefficient as for BSC, relating the step limit increase with a decrease of probability of failure.

Denote by $T(s)$ the probability that logarithm of the number of corrections to consider in a single moment is at most $s$. We know that $T(s)=0$ for $s<0$, but $T(0)>0$. Due to checking the redundancy bits, in each step the expected number of corrections is multiplied by $(1-p_d)$. For each bit lost, this number is additionally doubled. Finally, for $\nn=1$, we get functional equation (\ref{ereq}). Assuming as previously that asymptotically
$$1-T(s)\propto 2^{c_es}$$
for some $c_e\leq 0$, we get
$$2^{c_e s}=p_e 2^{c_e (s-\lg(\tilde{p_d})-1)}+(1-p_e)2^{c_e(s-\lg(\tilde{p_d}))}$$
for general $\nn$ this equation becomes:
\be \tilde{p_d}^{c_e}=(p_e 2^{-c_e}+ 1-p_e)^{\nn} \label{erasc}\ee
This $c_e$ works as the $c$ coefficient for BSC. It approaches 0 in the Shannon limit and effectively doubles for bidirectional correction.

\section*{Acknowledgment}

We would like to thank Professor Andrzej Pach, from the Department of Telecommunications,
AGH University of Science and Technology, for his guidance and help.

\bibliographystyle{plain}
\bibliography{error-correction}

\begin{thebibliography}{10}

\bibitem{correction-trees-sources}
C++ source code for the correction trees algorithm.
\newblock https://indect-project.eu/correction-trees/.

\bibitem{itpp}
The it++ library.
\newblock http://itpp.sourceforge.net/current/.
\newblock Version 4.2.

\bibitem{berrou}
C.~Berrou, A.~Glavieux, and P~Thitimajshima.
\newblock Near shannon limit error-correcting coding and decoding: Turbo-codes.
\newblock In {\em Proc. of IEEE International Conference on Communications},
  1993.

\bibitem{ans}
J.~Duda.
\newblock Asymmetric numeral systems.
\newblock arxiv: 0902.0271.

\bibitem{wolfram}
J.~Duda.
\newblock Correction trees, interactive wolfram mathematica demonstration.
\newblock http://demonstrations.wolfram.com/CorrectionTrees/.

\bibitem{conv}
P.~Eliaso.
\newblock “coding for noisy channels.
\newblock {\em ZRE Conv. Rept. Pt}, 4:37--47, 1955.

\bibitem{fano}
R.~M. Fano.
\newblock A heuristic discussion of probabilistic decoding.
\newblock {\em IEEE Transaction on Information Theory}, 9:64--73, 1963.

\bibitem{gallager}
R.~G. Gallager.
\newblock {\em Low-Density Parity-Check Codes}.
\newblock PhD thesis, Cambridge, MA, 1963.

\bibitem{cutoff}
I.~M. Jacobs and E.~R. Berlekamp.
\newblock A lower bound on to the distribution of computation for sequential
  decoding.
\newblock {\em IEEE Transactions on Information Theory}, IT-13:167--174, 1967.

\bibitem{mackay2003}
D.~J. MacKay.
\newblock {\em Information Theory, Inference, and Learning Algorithms}.
\newblock Cambridge University Press, 2003.

\bibitem{ldpc-encyclopedia}
David~J.C. MacKay.
\newblock Encyclopedia of sparse graph codes.
\newblock http://www.inference.phy.cam.ac.uk/mackay/codes/data.html.

\bibitem{mackay}
David~J.C. MacKay.
\newblock Good error correcting codes based on very sparse matrices.
\newblock {\em IEEE Transactions on Information Theory}, pages 399--431, March
  1999.

\end{thebibliography}
\end{document}